\documentclass{aa}  

\usepackage{graphicx}
\usepackage{txfonts}
\usepackage{color}
\usepackage{xcolor}
\usepackage{longtable}
\usepackage{txfonts}
\usepackage{rotating}
\usepackage{lscape}
\usepackage{booktabs}
\usepackage{multirow}
\usepackage{soul}

\begin{document}

   \title{Open clusters in APOGEE and GALAH}

   \subtitle{Combining \textit{Gaia} and ground-based spectroscopic surveys}

   \author{R. Carrera
          \inst{1}
          \and
          A. Bragaglia\inst{2}
          %\fnmsep\thanks{}
         \and T. Cantat-Gaudin\inst{3}
         \and A. Vallenari\inst{1}
         \and L. Balaguer-N\'u\~nez\inst{3}
         \and D. Bossini\inst{1}
        \and L. Casamiquela\inst{4}
         \and C. Jordi\inst{3}
         \and R. Sordo\inst{1}
       \and C. Soubiran\inst{4}}

\institute{INAF-Osservatorio Astronomico di Padova, vicolo dell'Osservatorio 5, 35122 Padova, Italy\\
              \email{jimenez.carrera@inaf.it}
         \and
    INAF-Osservatorio di Astrofisica e Scienza dello Spazio, via P. Gobetti 93/3, 40129 Bologna, Italy
    \and
Institut de Ci\`encies del Cosmos, Universitat de Barcelona (IEEC-UB), Mart\'i i Franqu\`es 1, E-08028 Barcelona, Spain  
\and
Laboratoire d'Astrophysique de Bordeaux, Univ. Bordeaux, CNRS, B18N, all\'ee Geoffroy Saint-Hilaire, F-33615 Pessac, France
    }

   \date{Received ; accepted }

% \abstract{}{}{}{}{} 
% 5 {} token are mandatory
 
  \abstract
   {Open clusters are ideal laboratories to investigate a variety of astrophysical topics, from the properties of the Galactic disk to stellar evolutionary models. Knowing their metallicity and possibly detailed chemical abundances is therefore important.  However, the number of systems with chemical abundances determined from high resolution spectroscopy is still small.}
  {To increase the number of open clusters with radial velocities and chemical abundances determined from high resolution spectroscopy we used publicly available catalogues of surveys
  in combination with \textit{Gaia} data.} 
   {Open cluster stars have been identified in the APOGEE and GALAH spectroscopic surveys by cross-matching their latest data releases with stars for which high-probability astrometric membership has been derived in many clusters on the basis of the \textit{Gaia} second data release. }
  {Radial velocities have been determined for 131 and 14 clusters from APOGEE and GALAH data, respectively. This is the first radial velocity determination from high resolution spectra for 16 systems. Iron abundances have been obtained for 90 and 14 systems from APOGEE and GALAH samples, respectively. To our knowledge 66 of these clusters (57 in APOGEE and 9 in GALAH) do not have previous determinations in the literature. For 90 and 7 clusters in the APOGEE and GALAH samples, respectively, we have also determined average abundances for Na, Mg, Al, Si, Ca, Cr, Mn, and Ni.
 %Finally, the obtained results have been used to investigate the %existence of radial and vertical chemical gradients in the Galactic %disk.
  }
   {}


   \maketitle
%
%________________________________________________________________

\section{Introduction}\label{sect:intro}

Open clusters (OCs) are groupings of between 10 and a few thousand stars that share chemo-dynamical features with a common birth-time and place. They are probably the only chemically homogeneous stellar populations \citep[e.g.][]{desilva2007,bovyOChomogenety} but see also \citet{liu2016}.
These systems play a fundamental role in our understanding of both individual and group stellar evolution allowing to investigate a variety of astrophysical topics such as initial mass function, initial binary fraction, the creation of blue stragglers, mass loss, or atomic diffusion among others. Thanks to the fact that OCs cover a wide range of ages and are found everywhere in the Galactic disk, they have been widely used to trace both the disk chemistry, e.g. disk metallicity gradient \citep[e.g.][]{friel02,donati15, jacobson16,netopil16,occaso2} and its evolution with time \citep[e.g.][]{andreuzzi2011,carrera&pancino2011}, and dynamics, e.g. individual orbits \citep[e.g.][]{cantat_gaudin2016,reddy2016} or radial migration \citep[e.g.][]{rovkar2008,anders2017}.

A detailed characterisation of the OCs chemical composition is necessary to fully exploit their capabilities to address the topics described above. High-resolution spectroscopy ($R\geq$20,000) is the most direct way to obtain chemical abundances; however, for some OCs these studies have been limited to the determination of the iron content, widely known as metallicity\footnote{There is some ambiguity in the use of the term of metallicity in the literature. Together with the iron abundance, typically denoted as [Fe/H], the term metallicity is also used to refer to the overall abundance of all elements heavier than helium, denoted as [M/H].}. Moreover, this kind of analysis has been performed for slightly more than 100 objects \citep[see e.g. the literature compilations by][]{carrera&pancino2011,yong12,heiter2014,donati15,netopil16}. They represent the 10\% of the about 3000 known OCs according to the updated versions of the two most used OCs compilations by \citet[][DAML]{dias2002} and \citet[][MWSC]{kharchenko13}. The real cluster population is still largely unknown; not only many of these 3000 objects need to be confirmed as real clusters \citep[see e.g. ][for objects that are likely not clusters]{galah_highlatOC,tristan18b}, but new clusters are being discovered thanks to surveys like the \textit{Gaia} mission (see later).

The \textit{Gaia} mission \citep{2016A&A...595A...1G} is carrying out a revolution in astronomy providing an unprecedented large volume of high quality positions, parallaxes and proper motions. This is supplemented by very high-accuracy all-sky photometric measurements. Additionally, for the brightest stars \textit{Gaia} is also providing  radial velocities \citep{sartoretti2018}
and in the future it will provide some information about their chemical composition \citep{bailer_jones2013}.

Complementing the limited spectroscopic capabilities of \textit{Gaia} is the motivation of the several ongoing and forthcoming ground-based high-resolution spectroscopic surveys providing radial velocities and chemical abundances for more than 20 chemical species. At the moment the \textit{Gaia}-ESO Survey \citep[GES][]{gilmore12,randich13} is the only high-resolution survey which has dedicated a significant fraction of time to target open clusters. \textit{Gaia}-ESO is providing an homogeneous set for about 80 clusters \citep[see e.g.][and references therein]{jacobson16,randich18} observed extensively (100-1000 stars targeted in each of them).  The other two high-resolution surveys with data published until now, APOGEE \citep[Apache Point Observatory Galactic Evolution Experiment;][]{majewski17} and GALAH \citep[Galactic Archaeology with HERMES;][]{desilva15}, do not have such a large and specific program on OCs, although they are targeting some of them, also for calibration purposes (see e.g. \citealt{occam2,kos18}). Their latest data releases include about 277,000 \citep{holtzman2018dr13dr14} and 340,000 \citep{buder18} stars, respectively.

This paper is the third of a series devoted to the study of OCs on the basis of \textit{Gaia}~DR2. In the first one, membership probabilities for OCs were derived from the \textit{Gaia}~DR2 astrometric solutions \citep{tristan18b}. In the second, the \textit{Gaia}~DR2 radial velocities were used to investigate the distribution of OCs in the 6D space \citep{soubiran18}. The goal of this paper is to search for cluster stars hidden in both the APOGEE and GALAH catalogues\footnote{We also searched the public \textit{Gaia}-ESO catalogue (DR3, see {\tt https://www.eso.org/qi/}) for unrecognised cluster stars but found only one additional star in one cluster, so we did not proceed further.} in order to increase the number of OCs with radial velocities and chemical abundances derived from high resolution spectroscopy. To do so, we use the astrometric membership probabilities obtained by \citet{tristan18b}. 

This paper is organized as follows. The observational material utilized in the paper is described in Sect.~\ref{sect:data}. The radial velocities are discussed in Sect.~\ref{sect:rv}. The iron and other elements abundances are presented in Sect.~\ref{sect:fe} and \ref{sect:other}, respectively. An example of the usefulness of the results obtained in previous sections to investigate the radial and vertical chemical distribution of OCs in the Galactic disk is shown in Sect.\ref{sect:trends}. Finally, the main conclusions of this paper are discussed in Sect.~\ref{sect:conclusion}. 

\section{The Data}\label{sect:data}

The \textit{Gaia}~DR2 provides 5-parameter astrometric solution \citep[positions, proper motions $\mu_{\alpha*}$, $\mu_{\delta}$, and parallaxes $\varpi$;][]{lindegren2018} and magnitudes in three photometric bands \citep[$G$, $G_{BP}$ and $G_{RP}$;][]{evans2018} for more than 1.3 billion sources \citep{gaia_gdr2}, plus radial velocities (RV) for more than 7 million stars \citep{katz18}. On the basis of \textit{Gaia}~DR2, \citet{tristan18b} determined membership probabilities for stars in 1229 OCs,
60 of which are new clusters serendipitously discovered in the fields analysed. Because of the large uncertainties of the proper motion and parallax determinations for faint objects, the analysis was limited to stars with $G\leq$ 18, corresponding to a typical uncertainty of 0.3 mas\,yr$^{-1}$ and 0.15\,mas in proper motion and parallax, respectively. To assign the membership probabilities, $p$, they used the unsupervised photometric membership assignment in stellar clusters (UPMASK) developed by \citet{2014A&A...561A..57K}. We refer the reader to that paper for details on how the probabilities are assigned.

\subsection{APOGEE}

In the framework of the third and fourth phases of the Sloan Digital Sky Survey \citep{2011AJ....142...72E,2017AJ....154...28B}, APOGEE \citep{majewski17} obtained  $R\sim$22,500 spectra in the infrared $H$-band, 1.5-1.7 $\mu$m. The fourteenth Data Release \citep[DR14,][]{2018ApJS..235...42A,holtzman2018dr13dr14} includes about 277,000 stars and provides RVs with a typical uncertainty of $\sim$0.1 km\,s$^{-1}$ \citep{apogeepipeline}. Because APOGEE tries to observe each star at least three times, the RV uncertainty, called $RV\_scatter$ and defined as the scatter among the individual RV determinations, provides a possible indication of stellar binarity. Stellar parameters and abundances for 19 chemical species are determined with the APOGEE stellar parameter and chemical abundance pipeline \citep[ASPCAP;][]{aspcap}. Briefly, ASPCAP works in two steps: it first determines stellar parameters using a global fit over the entire spectral range by comparing the observed spectrum with a grid of synthetic spectra, and then it fits sequentially for individual elemental abundances over limited spectral windows using the initially derived parameters. 

\begin{table}
\centering
\caption{Number of stars with a membership probability above a
given cut, and the corresponding number of OCs with at least one star.\label{tab_prob_sel}}
\setlength{\tabcolsep}{0.7mm}
\begin{tabular}{lcc}
\hline
$p$ & Nr Stars & Nr OCs\\
\hline
$\geq$0.1 & 1638 & 175\\
$\geq$0.2 & 1559 & 164\\
$\geq$0.3 & 1494 & 152\\
$\geq$0.4 & 1447 & 138\\
$\geq$0.5 & 1406 & 131\\
$\geq$0.6 & 1370 & 129\\
$\geq$0.7 & 1315 & 124\\
$\geq$0.8 & 1222 & 119\\
$\geq$0.9 & 1082 & 108\\
$=$1.0 & 852 & 84\\
\hline
\end{tabular}
\end{table}

APOGEE has observed a few OCs to serve as calibrators \citep[see][]{holtzman2018dr13dr14}. Other OC stars have been observed in the framework of the Open Cluster Chemical Abundances and Mapping (OCCAM) survey \citep{occam1,occam2} when the clusters were in the field of view of a main survey pointing. Finally, there may be also cluster stars observed by chance among the survey targets. The latter is the main goal of this paper.

We cross-matched the \textit{Gaia}~DR2 high probability OC members with the whole APOGEE DR14 dataset. We excluded those objects flagged in {\em STARFLAG} as having: many bad pixels ($\geq 40\%$), low signal-to-noise ratio ($\leq 50$ per half-resolution element), or potentially binary stars with significant RV variation among visits ($RV\_scatter\geq$5 km\,s$^{-1}$). We rejected also those objects that are clearly out of the cluster sequences, which usually have low probabilities, $p\leq$0.6. Finally, a dozen of stars were rejected because they have been reported as non cluster members on basis of their radial velocities in the literature. This step rejects one cluster, Berkeley~44, where the observed star has {\bf an} astrometric membership of 0.6 but it is a field object according to \citet{hayesfriel2014}. In fact the RV of this star is quite different from the mean value derived for this cluster in the literature.

\citet{tristan18b} provide discrete astrometric probabilities $p$ for each star to belong to its parent cluster. It takes values between 0.1, least likely, and 1.0, most likely, with a step of 0.1. The derived average RV and chemical abundances can significantly change as a function of the probability threshold used to select the most probable members. Moreover, the total number of OCs for which a mean RV, and also chemical composition, can be computed also depends on the adopted probability cut, as shown in Table~\ref{tab_prob_sel}. Although using a low probability cut can add stars and clusters to the analysis, it also increases the dispersion of the derived values since some low probability members are not real members. It is necessary, therefore, to find an optimal selection threshold. To do so, the strategy developed in \citet{soubiran18} has been followed computing the average RV for the 30 clusters with 4 or more stars with $p=$1  using different probability cuts. The mean RV values have been obtained using

\begin{equation}\label{eq:meanrv}
RV = \frac{\sum_i RV_i \times g_i}{\sum_i g_i}
\end{equation}
where $RV_i$ is the individual RV derived by APOGEE with the weight $g_i$ defined as $g_i=1/(RV\_scatter_{i})^2$, where, $RV\_scatter_{i}$ is the radial velocity scatter provided by APOGEE.

\begin{figure}
\centering\includegraphics[scale=0.35]{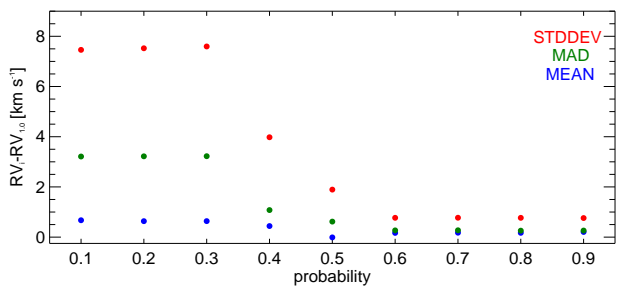}
\caption{Run of standard deviation (STDDEV, red), median absolute deviation (MAD, green), and mean (blue) of the difference between the mean RV obtained using different probability cuts and those obtained only for stars with \textit{p}$=$1 for the 30 clusters with 4 or more stars with the highest probability.}
\label{fig_test_prob}
\end{figure}

In Fig.~\ref{fig_test_prob} we have plotted the standard deviation (STDDEV, red), median absolute deviation (MAD; green), and mean (blue) of the difference between the mean RV obtained using different probability cuts, RV$_i$, and the value obtained using only stars with \textit{p}=1, RV$_{1.0}$. The mean of the difference does not change significantly for the different cuts, but we see a decrease at $p>0.5$. The MAD is almost flat for \textit{p}$\geq$0.4 and it differs significantly between \textit{p}=0.3 and \textit{p}=0.4. The same trend is observed in the standard deviation but in this case the main increase is observed between \textit{p}=0.4 and \textit{p}=0.5. As a result of this analysis we limit our analysis to those stars with \textit{p}$\geq$0.5. We note that \citet{soubiran18} considered members with probabilities \textit{p}$\geq$0.4 based on other reference values from the  literature.

\begin{figure*}
\centering\includegraphics[scale=0.9]{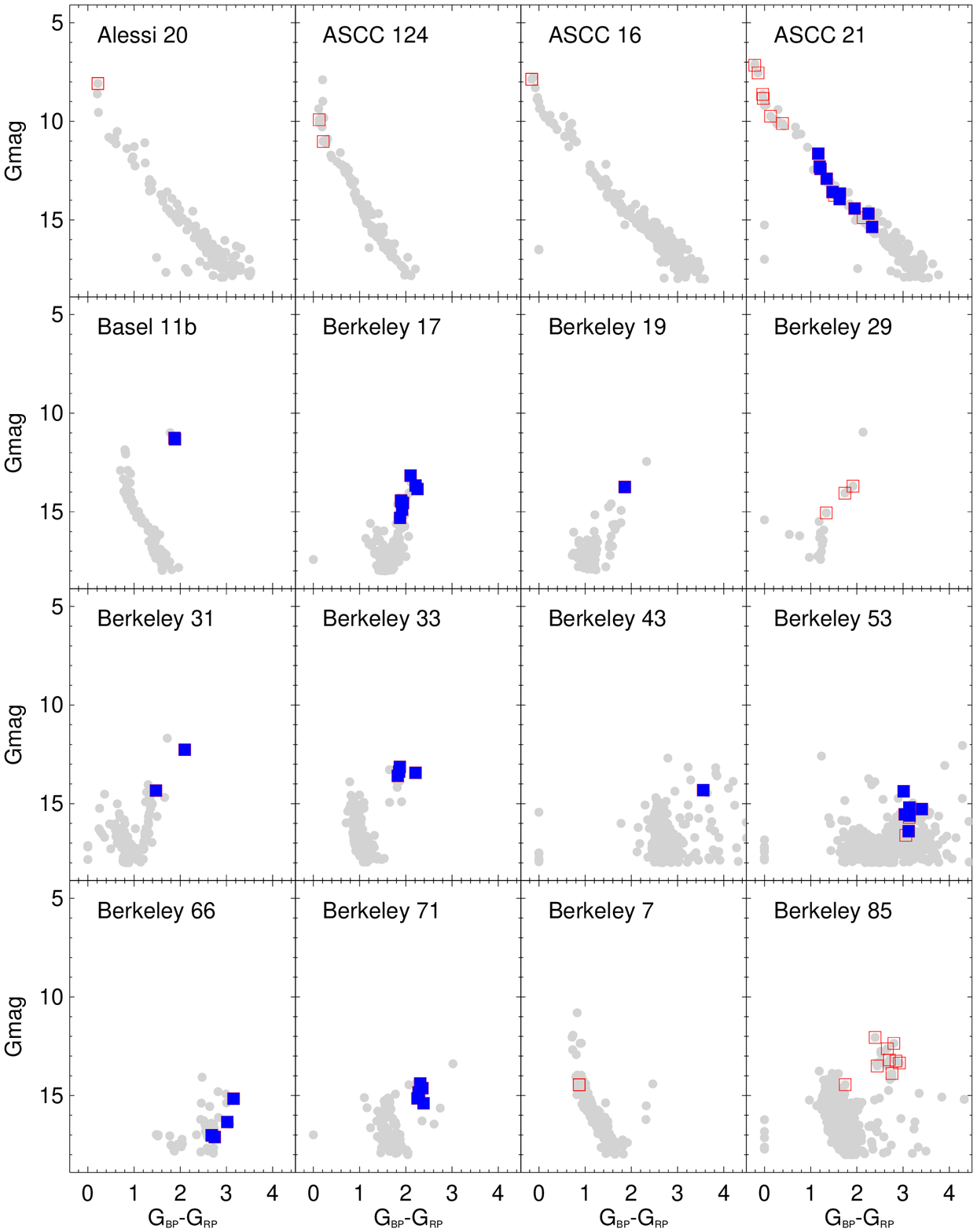}
\caption{\textit{Gaia}~DR2 colour-magnitude diagrams for OCs with stars in common with APOGEE DR14. Grey circles are \textit{Gaia}~DR2 stars with a membership probability above 0.5. Open red squares are stars in common with APOGEE~DR14 used in the RV determination (see text for details). Filled blue squares are objects used in the [Fe/H] analysis.}
\label{fig_dcm_apogee}
\end{figure*}

In total we found 1406 stars with $p\geq$0.5 belonging to 131 open clusters in common between APOGEE DR14 and \textit{Gaia}-DR2 \citep{tristan18b}. These 131 systems are listed in Table~\ref{tab_apogee_rv}. A few examples of colour-magnitude diagrams (CMD) of these clusters are shown in Fig.~\ref{fig_dcm_apogee}. The individual stars are listed in Table~\ref{A1}.

\begin{table*}
\centering
\caption{The 131 open clusters in common between APOGEE DR14 and \citet{tristan18b}.\label{tab_apogee_rv}}
\setlength{\tabcolsep}{0.45mm}
\begin{tabular}{lc|cccc|ccc|ccc|cccc|ccc|c}
\hline
Cluster & Star & RV & $\sigma_{RV}$ & e$_{RV}$ & Nr & RV$_{lit}$ & $\sigma_{RV,lit}$ & Nr$_{lit}$ & RV$_{GDR2}$\tablefootmark{b} & $\sigma_{GDR2}$ & Nr$_{GDR2}$ & [Fe/H] & $\sigma_{[Fe/H]}$ & e$_{[Fe/H]}$ & Nr & [Fe/H]$_{lit}$ & $\sigma_{[Fe/H],lit}$ & Nr$_{lit}$ & Ref.\\
 & type\tablefootmark{a}  &\multicolumn{3}{c}{(km~s$^{-1}$)} & & \multicolumn{2}{c}{(km~s$^{-1}$)}&  & \multicolumn{2}{c}{(km~s$^{-1}$)}&  & \multicolumn{3}{c}{(dex)} & & \multicolumn{2}{c}{(dex)}& & \\
\hline
  Alessi~20 & MS & -14.89 & & 0.37 & 1 & -11.5 & 0.01 & 2 & -5.04 & 3.3 & 7 &  &  &  &  &  &  &  &  1\\
  ASCC~124 & MS & -23.35 & 0.54 & 0.38 & 2 &  &  &  &  &  &  &  &  &  &  &  &  &  & \\
  ASCC~16 & MS & 17.41 &  & 0.62 & 1 &  &  &  & 23.18 & 3.4 & 15 &  &  &  &  &  &  &  & \\
  ASCC~21 & MS & 16.30 & 5.66 & 1.30 & 19 & 18.57 & 2.12 & 9 & 18.7 & 4.45 & 9 & 0.01 & 0.09 & 0.03 & 10 & & & & 2\\
  Basel~11b & RGB & 2.68 & 0.06 & 0.04 & 2 &  &  &  & 3.43 & 1.46 & 3 & 0.014 & 0.05 & 0.04 & 2 &  &  &  &  \\
  Berkeley~17 & RGB & -73.34 & 0.41 & 0.13 & 9 & -73.4 & 0.4 & 7 & -71.95 & 1.77 & 7 & -0.10 & 0.04 & 0.01 & 9 & -0.11 & 0.03 & 7 & 3\\
  Berkeley~19 & RGB & 17.44 & 0.0 & 0.08 & 1 &  &  &  & 17.65 & 0.42 & 1 & -0.22 & & 0.01 & 1 &  &  &  & \\
  Berkeley~29 & RGB & 25.27 & 0.53 & 0.30 & 3 & 24.8 & 1.13 & 11 & 50.58 & 0.94 & 1 &  &  &  &  &  &  &  & 1\\
  Berkeley~31 & RGB & 57.87 & 0.65 & 0.46 & 2 & 61.0 & 3.75 & 17 &  &  &  & -0.305 & 0.037 & 0.026 & 2 & -0.31 & 0.06 & 2 & 1,12\\
  Berkeley~33 & RGB & 77.80 & 0.56 & 0.28 & 4 & 76.6 & 0.5 & 5 & 78.82 & 1.12 & 4 & -0.23 & 0.11 & 0.05 & 4 &  & &  & 2\\
  Berkeley~43 & RGB & 29.712 &  & 0.136 & 1 &  &  &  & 29.15 & 1.42 & 9 & 0.00 & & 0.01 & 1 &  & &  & \\
  Berkeley~53 & RGB & -35.77 & 0.82 & 0.29 & 8 & -36.3 & 0.5 & 4 & -34.9 & 1.81 & 7 & -0.02 & 0.03 & 0.01 & 6 & 0.00 & 0.02 & 5 & 3\\
\hline
\end{tabular}
\tablebib{(1)~\citet{kharchenko13};
(2)~\citet{dias2002}; (3)~\citet{occam2}; (4)~\citet{occaso1};
(5)~\citet{carrera2017king1}; (6)~\citet{donati2015tr5}; (7)~\citet{magrini17}; (8)~\citet{schiappacasse2018}; (9)~\citet{blanco_cuaresma2015}; (10)~\citet{heiter2014}; (11)~\citet{zavcs2011}; (12)~\citet{friel2010}; (13)~\citet{monroe2010}.
}
\tablefoot{
\tablefoottext{a}{Type of stars used in the analysis: (RGB) red giant branch, (MS) main-sequence, or both.}.
\tablefoottext{b}{From \citet{soubiran18}.}
The entire version will be available on-line.}
\end{table*}

\subsection{GALAH}

The GALAH survey \citep{desilva15,martell2017} is a Large Observing Program using the High Efficiency and Resolution Multi-Element Spectrograph (HERMES, \citealt{barden2010,sheinis2015}). HERMES provides simultaneous spectra of up to 392 objects with a resolution power of 28,000 in four wavelength bands: 4713-4903\,\AA~(blue arm), 5648-5873\,\AA~(green arm), 6478-6737\,\AA~(red arm), and 7585-7887\,\AA~(infrared arm). The GALAH second Data Release \citep[DR2][]{buder18} includes about 340,000 stars. Radial velocities and their uncertainties in GALAH are computed by cross-correlating the observed spectra with 15 synthetic AMBRE spectra \citep{ambre}. The typical RV uncertainty in GALAH DR2 is $\sim$0.1\,km\,s$^{-1}$ 
\citep{zwitter2018}. GALAH chemical analysis is performed in two steps \citep[see][for details]{buder18}. Briefly, the stellar parameters and abundances of a training set of about 10,500 stars are first found by spectral synthesis with the Spectroscopy Made Easy code \citep[SME,][]{valenti1996,piskunov2017}. The obtained results are then used to train the {\sl The Cannon} \citep{ness2015} data-driven algorithm to find stellar parameters and abundances for the whole GALAH sample.

According to \cite{buder18}, open clusters are not part of the fields already released by GALAH but are observed by several separate programmes with HERMES, i.e. with the same instrument. Some OCs were  used in \citet{buder18} as a test of the GALAH results (see later for further details). \citet{kos18} combined \textit{Gaia}~DR2 and GALAH to study five candidate low-density, high-latitude clusters, finding that only one of them, NGC~1901, can be considered a cluster, while the others are only chance projections of stars. Similar results have been found by \citet{tristan18b} for many high-latitude candidate clusters.

While GALAH and the linked private projects are targeting OCs on purpose, there may be also cluster stars observed serendipitously and we looked for them. After cross-matching the \citet{tristan18b} high-probability members ($p\ge0.5$) with GALAH DR2 we found 122 stars in 14 OCs. We list in Table~\ref{A2} parameters from \textit{Gaia} ($G$, $G_{BP}$-$G_{RP}$, $\varpi$, $\mu_{\alpha*}$, $\mu_{\delta}$) and GALAH (RV, T$_{\rm eff}$, $\log g$, $v\sin i$, [Fe/H]) for the individual stars. Among them, we selected those stars without problems during {\sl The Cannon} analysis, labelled as {\em flag\_cannon=0}, or that need only some extrapolation, {\em flag\_cannon=1} \citep[see][for more details]{buder18}. After applying these constrains we are left with 82 stars in 14 clusters. 

Since the majority of the clusters are young, the stars targeted are mostly on the main-sequence (MS; there are giants only in NGC~2243 and NGC~2548).
Furthermore, they are often of early spectral types and may show high rotational velocities. Their RV determination is then less reliable (the same is valid for metallicity, see Sect. 4.2) and we tried to keep only the stars with $v\sin i$ values lower than 20 km~s$^{-1}$. In addition,  candidate members according to astrometry show discrepant RVs in some clusters (see Table~\ref{A2}). To select only the highest probability cluster members based both on astrometry and RV, we used the average cluster RVs determined by \citet{soubiran18} using \textit{Gaia}~DR2 data (at least 4 stars were sampled in all these OCs). This affects only five clusters: ASCC~21, Alessi~24, Alessi-Teutsch~12, NGC~5640, and Turner~5. One discrepant radial velocity star have been discarded in each of them.

\begin{figure*}
\centering\includegraphics[scale=0.9]{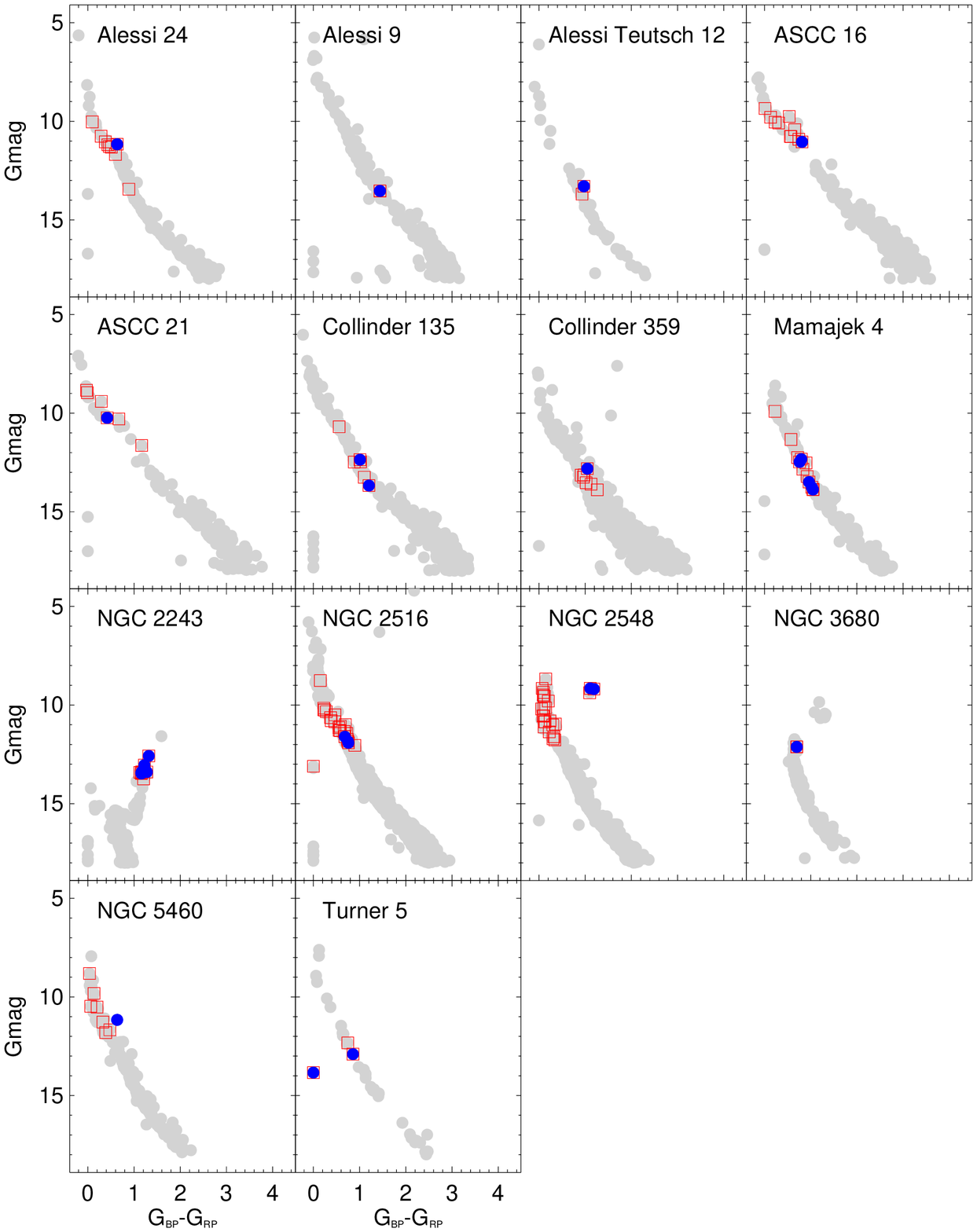}
\caption{As in Fig.~\ref{fig_dcm_apogee} for the 14 clusters in the GALAH sample. Note that one star in NGC~2516 and one in Turner~5 have no colour information and were arbitrarily plotted at $G_{BP}$-$G_{RP}$=0.} %Black open squares are all the stars observed by GALAH in the field of view of each cluster. Blue filled circles are the objects used in the analysis with {\em flag\_cannon} equal to 0 or 1.}
\label{dataG}
\end{figure*}

After all this weeding we ended up with 29 stars in 14 clusters; in half of the cases only one star survived the selections. The properties of these stars are summarised in Table~\ref{selected_G},  Fig.~\ref{dataG} shows the CMD of the 14 OCs and the stars used in our analysis indicated, while Table~\ref{oc_GALAH} lists the 14 OCs and their mean RV and metallicity.

\begin{table*}
\centering
\caption{Properties of the 14 open clusters in common between GALAH DR2 and \citet{tristan18b}.}\label{oc_GALAH}
\setlength{\tabcolsep}{0.4mm}
\begin{tabular}{lc|ccccc|ccc|cccc|cccc}
\hline
Cluster & Star & RV & $\sigma_{RV}$ & e$_{RV}$ & $\left\langle{\rm v}\sin i\right\rangle$\tablefootmark{a}star & Nr & RV$_{GDR2}$\tablefootmark{b} & $\sigma_{GDR2}$ & Nr$_{GDR2}$ & [Fe/H] & $\sigma_{[Fe/H]}$ & e$_{[Fe/H]}$ & Nr & [Fe/H]$_{lit}$ & $\sigma_{[Fe/H],lit}$ & Nr$_{lit}$ & Ref.\\
&type    &\multicolumn{4}{c}{(km~s$^{-1}$)} & & \multicolumn{2}{c}{(km~s$^{-1}$)}&  & \multicolumn{3}{c}{(dex)} & & \multicolumn{2}{c}{(dex)}& & \\
\hline
Alessi~24 & MS & 10.44 & & 0.11 & 11.63 & 1 & 12.32 & 2.24 & 14 & -0.13 & & 0.07 & 1 & & & & \\
Alessi~9  & MS & -5.64 & & 0.11 &  5.70 & 1 & -6.44 & 1.14 & 39 & -0.06 & & 0.07 & 1 & & & & \\
Alessi-Teutsch~12 & MS & -11.00 & & 0.12 & 5.35 & 1 & -5.88 & 3.54 & 6 & 0.40 & & 0.08 & 1 & & & & \\
ASCC~16 & MS & 22.01 & & 0.17 & 38.82\tablefootmark{c} & 1 & 23.18 & 3.40 & 15 & -0.08\tablefootmark{d} & & 0.06 & 1 & & & & \\
ASCC~21 & MS & 20.05 & & 1.06 & 19.99\tablefootmark{c} & 1 & 18.70 & 4.45 & 9 & -0.13\tablefootmark{d} & & 0.08 & 1 & & & & \\
Collinder~135& MS & 17.08 & 0.96 & 0.68 & 18.31 & 2 & 16.03 & 2.21 & 51 & -0.09 & 0.03 & 0.02 & 2 & & & & \\
Collinder~359 & MS & 8.30 & & 1.79 & 57.53\tablefootmark{c} & 1 & 5.28 & 3.25 & 12 & -0.66\tablefootmark{d} & & 0.08 & 1 & & & & \\
Mamajek~4 & MS & -28.09 & 2.19 & 0.98 & 6.95 & 5 & -26.32 & 3.17 & 34 & 0.09 & 0.17 & 0.08 & 5 & & & & \\
NGC~2243 & RGB & 59.32 & 0.57 & 0.23 & 7.36 & 6 & 59.63 & 1.06 & 4 & -0.31 & 0.05 & 0.02 & 6 & -0.43 & 0.04 & 16 & 1 \\
NGC~2516& MS & 26.01 & 1.40 & 0.81 & 34.10\tablefootmark{c} & 3 & 23.85 & 2.01 & 132 & -0.26\tablefootmark{d} & 0.05 & 0.03 & 3 & 0.06 & 0.05 & 13 & 1 \\
NGC~2548 & RGB & 8.45 & 0.40 & 0.28 & 5.20 & 2 & 8.85 & 1.08 & 14 & 0.16 & 0.01 & 0.04 & 2 & & & & \\
NGC~3680 & MS & 2.98 & & 1.99 & 33.54\tablefootmark{c} & 1 & 1.74 & 1.36 & 30 & -0.26\tablefootmark{d} & & 0.07 & 1 & -0.01 & 0.06 & 10 & 2\\
NGC~5460 & MS & -5.16 & 0.31 & 0.22 & 23.79\tablefootmark{c} & 2 & -4.61 & 2.77 & 5 & -0.32\tablefootmark{d} & 0.15 & 0.11 & 2 & & & & \\
Turner~5 & MS & -2.83 & 0.09 & 0.07 & 11.77 & 2 & -3.49 & 1.93 & 6 & 0.02 & 0.17 & 0.12 & 2 & & & & \\
\hline
\end{tabular}
\tablebib{(1)~\citet{magrini17}; (2) \citet{netopil16}}
\tablefoot{
\tablefoottext{a}{Mean of the individual values for those clusters with more than one stars.}
\tablefoottext{b}{From \citet{soubiran18}}
\tablefoottext{c}{Values for clusters where $\left\langle{\rm v}\sin i\right\rangle > 20$  km~s$^{-1}$ are considered uncertain.}
\tablefoottext{d}{[Fe/H] uncertain because $\left\langle{\rm v}\sin i\right\rangle > 20$  km~s$^{-1}$.}
}
\end{table*}

\section{Radial Velocities}\label{sect:rv}

\subsection{APOGEE}

The mean RV for each cluster has been computed using equation~\ref{eq:meanrv} described above. The internal velocity dispersion is derived as

\begin{equation}\label{eq:s_rv}
\sigma_{RV}=\sqrt{\frac{n}{n-1}\times\frac{\sum_i g_i\times(RV_i-RV)^2}{\sum_i g_i}}
\end{equation}
with an uncertainty of
\begin{equation}\label{eq:e_rv}
e_{RV}=\frac{\sigma_{RV}}{\sqrt{n}}
\end{equation}

For those clusters with only one star sampled we did not compute $\sigma_{RV}$ and we assumed $e_{RV}=RV\_scatter_{i}$. The obtained values are listed in Table~\ref{tab_apogee_rv}. In total we have determined mean RV for 131 clusters. For 78 of them, about 65\% of the total, this RV determination is based on less than 4 stars and the values have to be taken with caution. For the other 53 systems the RV is determined from 4 stars or more. In principle these values are more reliable except if they show a large $\sigma_{RV}$. Large $\sigma_{RV}$ values can be due to undetected binaries but also by residual field stars contamination. 

For the majority of the OCs the RV determination is based on either giant or MS stars. In general, the systems whose RVs have been determined from MS stars have larger $\sigma_{RV}$ values because of the larger RV uncertainties for these stars. There are 9 clusters with both kind of stars sampled: Berkeley~71, Berkeley~85, Berkeley~9, King~7, NGC~1664, NGC~1857, NGC~2682, NGC~6811, and NGC~7782. Except for NGC~1664, at least 5 stars have been observed in each of them. There are not significant differences in the mean RV values if we use only MS stars or giants.

\begin{figure}
\centering
\includegraphics[scale=0.35]{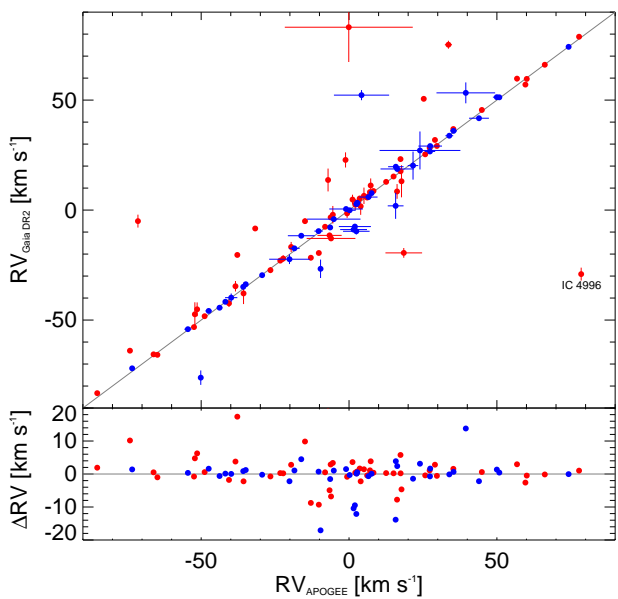}
\caption{Comparison between RV derived from APOGEE DR14 and \textit{Gaia}~DR2. Blue points are clusters with 4 or more stars while red points are systems with less then 4 stars.}
\label{fig_rv_apogeegdr2}
\end{figure}

There are 104 clusters in common with the recent work by \citet{soubiran18}. They derived 
mean RVs for 861 OCs based on the \textit{Gaia}~DR2 catalogue and the RVs obtained with the \textit{Gaia} RVS instrument, using a pre-selection done on our same astrometric membership probabilities \citep{tristan18b}. The top panel of Fig.~\ref{fig_rv_apogeegdr2} shows the comparison between the RV for the 104 clusters in common. The RV differences, $\Delta RV$, defined as $RV_{\textit{Gaia}~DR2}-RV_{APOGEE}$, are shown in the bottom panel of Fig.~\ref{fig_rv_apogeegdr2}. In general, there is a very good agreement between both samples for most of the clusters in spite of the \textit{Gaia}~DR2 typical uncertainties being larger than 2.5\,km\,s$^{-1}$. The median difference between \textit{Gaia}~DR2 and our APOGEE sample is 0.4\,km\,s$^{-1}$ with a median absolute deviation of 3.2\,km\,s$^{-1}$. However, there are a few cases that show significantly different RVs. All these clusters have only a few sampled stars either in our APOGEE sample or in the \textit{Gaia}~DR2 sample. For example, in the case of the system with the largest difference, IC~4996, the values in both samples have been obtained for only one candidate member. The RV from the star observed by APOGEE is 78.5$\pm$0.3\,km\,s$^{-1}$, while in \textit{Gaia}~DR2, from a different star, the obtained value is -29.1$\pm$2.9\,km\,s$^{-1}$. None of them is similar to the value of -2.5$\pm$5.7\,km\,s$^{-1}$ based on 4 stars cited in \cite{kharchenko13}, or to the value of $-12\pm5$ km~s$^{-1}$ found from pre-main-sequence stars by \cite{delgado99}. More data are required in cases such as this.

\begin{figure}
\centering
\includegraphics[scale=0.35]{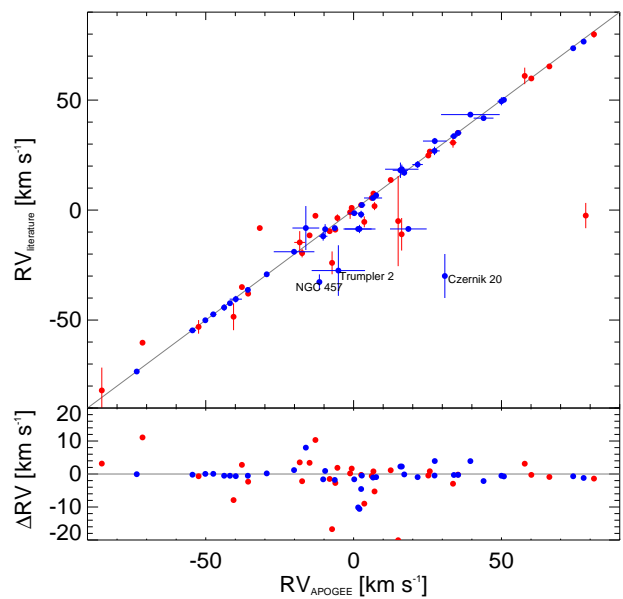}
\caption{Comparison between RV derived from APOGEE DR14 and literature. Blue and red points have the same meaning as in Fig~\ref{fig_rv_apogeegdr2}.}
\label{fig_rv_apogeelit}
\end{figure}

In addition to the \citet{soubiran18} work, we have compiled other RV determinations in the literature using the updated version of the DAML and MWSC catalogues as starting point and adding recent determinations available in the literature. In total we have compiled RVs for 75 of the 131 OCs in our sample (see Table~\ref{tab_apogee_rv} where also references are indicated). The comparison between the RV values obtained in this work and the literature is shown in the top panel of Fig.~\ref{fig_rv_apogeelit}, while the bottom panel shows the behaviour of the differences between them. As before, the largest discrepancies are generally observed for those clusters with less than 4 stars sampled (red points). In fact, the literature determinations for these clusters are also based on 4 objects or less, with the exception of NGC~457. In Czernik~20 and Trumpler~2 we have 7 and 5 stars, respectively. For Trumpler~2 we found a large $\sigma_{RV}$ which implies that our RV determination is not very reliable (see below). In the case of Czernik~20 the literature value has been obtained from a single star with an uncertainty of 10\,km\,s$^{-1}$. This value is much larger than the internal dispersion found in our work from 7 stars. For this reason we consider our RV determination more reliable.

To our knowledge this is the first RV determination for 16 clusters: ASCC~124, Berkeley~7, Berkeley~98, Czernik~18, Dolidze~3, FSR~0826, FSR~0941, Kronberger~57, L~1641S, NGC~1579, NGC~6469, Ruprecht~148, Stock~4, Teutsch~1, Teutsch~12, and Tombaugh~4. In the case of L~1641S our RV determination is based on 43 stars with an internal dispersion $\sigma_{RV}\sim$2\,km\,s$^{-1}$. Two clusters, Berkeley~98 and Teutsch~12, have 5 potential members with a $\sigma_{RV}$ of 3.3 and 2.3\,km\,s$^{-1}$, respectively. Due to the number of objects used in these three systems to derive their mean RV and given their small $\sigma_{RV}$, we believe that our determinations are reliable. For the remaining clusters the RV determinations are based on one star only, with the exception of ASCC~124 and Czernik~18 with two and three stars, respectively. As commented before, their RV are less reliable.

Eight systems have $\sigma_{RV}$ values larger than 6\,km\,s$^{-1}$. The largest internal dispersion is obtained for NGC~366 with $\sigma_{RV}\sim$21\,km\,s$^{-1}$. This value has been obtained from 4 stars with $p\geq$0.9 but with very different individual RV. In the case of NGC~2183 we found $\sigma_{RV}\sim$13\,km\,s$^{-1}$. This value is obtained from 5 stars with RVs between 7.4 and 43.1 \,km\,s$^{-1}$. However the two stars with the highest priorities, $p$=1 and 0.8, have RV of 28.9$\pm$0.2 and 25.4$\pm$0.1\,km\,s$^{-1}$, respectively. All the stars in NGC~2183 have only one APOGEE visit and therefore the RV determinations are more uncertain. The remaining six clusters (Koposov~36, NGC~2304, NGC~7086, Rosland~6, Trumpler~2, and Trumpler~3) have $\sigma_{RV}$ between 6 and 10\,km\,s$^{-1}$. The number of potential members sampled by APOGEE in these systems is between 4 and 8 stars. Almost all the stars in these clusters have higher membership probabilities, $p\geq$0.8. Since all of them are main-sequence stars, their individual RV determination can be affected by the large rotational velocities. Three of these clusters have determination of their RV in the literature (see Table~\ref{tab_apogee_rv}). In spite of the large RV dispersion observed in two of these clusters, NGC~2304 and NGC~7086, we find a good agreement with the values listed in the literature \citep{wu2009,kharchenko13}.

\subsection{GALAH}

\begin{figure}
\centering
\includegraphics[scale=0.35]{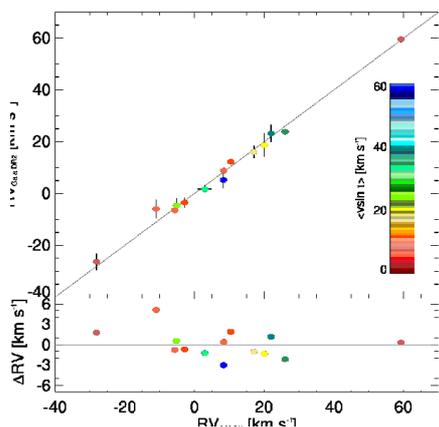}
\caption{Comparison between RV derived from GALAH DR2 and \textit{Gaia}~DR2, colour-coded according to $\left\langle{\rm v}\sin i\right\rangle$.}
\label{rvgalahgaia}
\end{figure}

 The same procedure followed above for APOGEE has been used to derive the mean RV, internal velocity dispersion, and uncertainty for the 14 OCs in the GALAH sample. The obtained values are listed in Table~\ref{oc_GALAH}.
 
Radial velocities for all the clusters in the GALAH sample have been determined previously by \citet{soubiran18} from \textit{Gaia}~DR2. The comparison between the RVs measured by GALAH and those by \textit{Gaia}~DR2 for the 14 clusters is shown in Fig.~\ref{rvgalahgaia}. The median difference between \textit{Gaia}~DR2 and GALAH is 0.3\,km\,s$^{-1}$ with a median absolute deviation of 1.5\,km~s$^{-1}$. The RVs are in reasonable agreement taken into account that the individual RV uncertainties in the \textit{Gaia}~DR2 are on average of 2.5\,km\,s$^{-1}$. The largest differences for some clusters (e.g. Collinder~359 and NGC~2517) are explained by their large average $\left\langle{\rm v}\sin i\right\rangle$ values. 

%\begin{figure}
%\centering
%\includegraphics[bb=20 50 550 560, clip, scale=0.45]{rv_fe_ave_galah.eps}
%\caption{Average metallicity and RV for the 14 clusters in GALAH DR2.}
%\label{rv_met_G}
%\end{figure}

The clusters NGC~2243 and NGC~2516 have been observed also by \textit{Gaia}-ESO and their mean RV are +60.2 (rms 1.0)\,km~s$^{-1}$ and +23.6 (rms 0.8)\,km~s$^{-1}$, respectively \citep{jackson,magrini17}. For NGC~2243 we obtained 59.3\,km~s$^{-1}$ with $\sigma_{RV}$=0.6\,km~s$^{-1}$ from 6 giant stars. The values are in good agreement within the uncertainties. In the case of NGC~2516 we obtained 26.0\,km~s$^{-1}$ with $\sigma_{RV}$=1.4\,km~s$^{-1}$ from 3 MS stars. The difference between GALAH and \textit{Gaia}-ESO mean RV could be due to the fact that the three stars in the GALAH sample have v$\sin i$ larger than 29\,km~s$^{-1}$.

Four of the GALAH clusters are also among the APOGEE systems discussed in previous section. These clusters are ASCC~16, ASCC~21, Collinder~359, and NGC~2243. In the case of NGC~2243 there is a good agreement between the values obtained from both samples in spite of the APOGEE value being based on only one star. For the other three clusters the differences between GALAH and APOGEE are of the order of $\pm$4\,km\,s$^{-1}$. This is not a large difference taken into account that all the stars sampled in these clusters by GALAH have $v\sin i$ larger than 20\,km\,s$^{-1}$.

\section{Metallicity}\label{sect:fe}

\subsection{APOGEE}

Before deriving average metallicity we excluded the stars for which the ASPCAP pipeline is not able to find a proper solution (or not a solution at all) because they are outside of or close to the edges of the synthetic library used in the analysis, e.g. hot stars \citep[see][for details]{aspcap,holtzman2018dr13dr14}. In this case the stars are flagged in {\em ASPCAPFLAG} as {\em STAR\_BAD} or  {\em NO\_ASPCAP\_RESULT}, respectively. After rejecting these objecys, the sample is reduced to 862 stars belonging to 90 clusters. Most of the rejected stars are fast rotating or low gravity objects.

Together with the individual iron abundance [Fe/H] for each star, APOGEE DR14 provides the scaled-solar general metallicity, [M/H]. The former is obtained from individual Fe lines whereas the latter is determined as a fundamental atmospheric parameter at the same time as effective temperature, surface gravity, and microturbulent velocity. In this paper we focus on the individual iron abundances. 
In contrast with the RV case, the average [Fe/H] of each cluster has been obtained as the un-weighted mean of the individual iron abundances. We have computed also $\sigma_{[Fe/H]}$ as the un-weighted standard deviation and $e_{[Fe/H]}$ as $\frac{\sigma_{[Fe/H]}}{\sqrt{n}}$\footnote{Other alternatives have been checked because several clusters can be affected by contamination of non-members such as the weighted mean or a Montecarlo simulation. In the first case we used the individual metallicity uncertainties as weights. For the Montecarlo simulation half of the stars in a given cluster were selected randomly and computed their mean and standard deviation. This procedure was repeated 10$^3$ times and the cluster mean and $\sigma$ where obtained as the mean of the individual means and standard deviations. In any case the differences between the obtained values and those derived in the paper are {\bf no} larger than $\pm$0.02\,dex.}.
Again for those clusters with only one sampled star we did not compute the standard deviation, $\sigma_{[Fe/H]}$, and we assume the uncertainty $e_{[Fe/H]}$, as the uncertainty for this star, $\sigma_{[Fe/H],i}$. 

The obtained values are listed in Table~\ref{tab_apogee_rv}. For 32 systems our [Fe/H] determination is based on 4 stars or more. Except for Berkeley~33, the $\sigma_{[Fe/H]}$ is lower than 0.1 dex, the typical uncertainty in the APOGEE [Fe/H] determination. The [Fe/H] determination for the other 58 clusters is based on less than 4 stars and typically on only one object.

\begin{figure}
\centering
\includegraphics[scale=0.35]{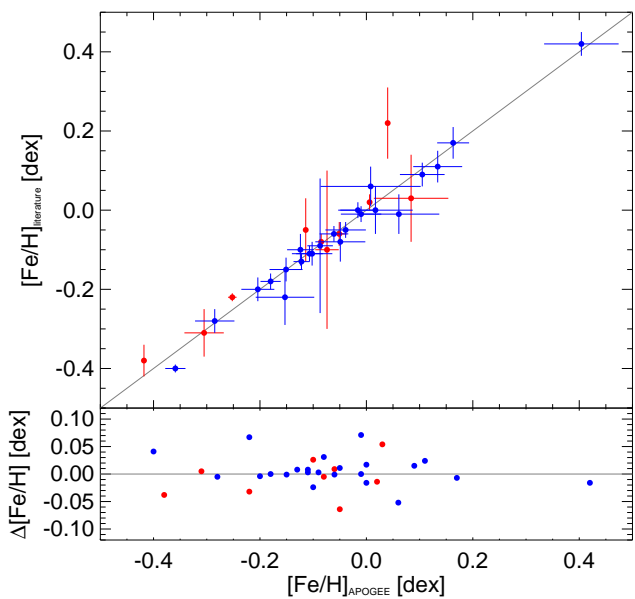}
\caption{Comparison between [Fe/H] derived from APOGEE DR14 and literature. Red and blue points are clusters with less and more than 4 stars, respectively.}
\label{fig_feh_apogeelit}
\end{figure}

To our knowledge there are previous determinations of iron abundances from high resolution spectra for a third of the total sample. The comparison between the values derived here and the literature (see Table~\ref{tab_apogee_rv} for references) is shown in Fig.~\ref{fig_feh_apogeelit}. In general, there is a very good agreement, with a median difference 0.00\,dex with a median absolute deviation of 0.02\,dex. This is not unexpected since several OCs have been used as reference to calibrate the whole APOGEE sample assuming values available in the literature \citep[see][for details]{holtzman2018dr13dr14}.

For the 57 clusters not studied previously only in 9 of them the metallicity determination is based on 4 or more stars. Two or three stars have been sampled for 16 of these systems. Finally, the metallicity determination for 32 previously unstudied clusters is based on a single star.

\subsection{GALAH}

In the case of GALAH the constraints applied initially also ensure that the iron abundance has been determined for all the stars in the sample. The average metallicities for the clusters in the GALAH sample has been derived using the same procedure as in the case of APOGEE. The obtained values are listed in Table~\ref{oc_GALAH}.

Only in two clusters (Mamajek~4 and NGC~2243) our analysis is based on more than 4 stars. The value obtained here for NGC~2243 is in good agreement with the result obtained by e.g. \citet{magrini17} and other literature sources. This cluster is also among the clusters studied in the previous section from APOGEE. Although the APOGEE analysis is based on a single star, the values are in agreement within the uncertainties.

Due to the early spectral type and high rotational velocity of the members, we judged as unreliable the metallicity determination for 6 clusters (ASCC~16, ASCC~21, Collinder~359, NGC~2516, NGC~3680, and NGC~5460). Moreover, their [Fe/H] values have been determined from 3 stars or less. One of these clusters, ASCC~21, has been also analysed from APOGEE data. The values obtained from APOGEE and GALAH samples show a difference of only 0.1\,dex, in spite of the large $v\sin i$ and high temperature of the only star in GALAH~DR2 for this cluster. Other two of these clusters, NGC~2516 and NGC~3680, have been studied previously in the literature  \citet[e.g.][]{magrini17} and \citet{netopil16}. In comparison with these works, the [Fe/H] values obtained here are about 0.25\,dex lower. \citet{buder18} discussed possible shortcomings of GALAH DR2 catalogue. They note that the double-step analysis is tailored on single, non-peculiar stars of F, G, and K spectral types and that some systematic trends may be present. In particular, they note difficulties for hot stars (i.e., hotter than late-F spectral type) because they have weaker metal lines, often rotate significantly, and are not present in the training set; all stars hotter than 7000\,K are then in extrapolation. Furthermore, there is a trend in the derived temperatures, with hotter stars showing lower temperatures than the comparison samples (e.g. the \textit{Gaia} Benchmark stars or stars for which the IRFM was available, their Fig.~14). In turn, this implies a trend towards lower metallicities. This is noticeable, for instance, in the T$_{\rm eff}$, $\log g$ diagrams for open clusters (their Fig.~19), with brighter MS stars showing systematically lower abundances. This is also what we found for NGC~2516 and NGC~3680.

In summary, the GALAH DR2 sample provides the first [Fe/H] determination for 10 clusters although with different degrees of reliability because of the large rotational velocities of some of the stars analysed.

\section{Other elements}\label{sect:other}

\subsection{APOGEE}

Although APOGEE DR14 provides abundances for 22 chemical elements, not all of them are completely reliable \citep[see][]{apogeeabundances,holtzman2018dr13dr14}. For this reason we limit our analysis to the elements that show small systematic differences in comparison with other literature samples according to \citet{apogeeabundances}. This includes $\alpha$-elements (Mg, Si, and Ca) and proton-capture elements (Na and Al) as well as elements of the iron-peak group (Cr, Mn and Ni). For each element we have excluded the stars flagged by ASPCAP with problems in the abundance determination. %: i.e. if the obtained value is close to the grid edge.

We have been able to determine abundances for all the 90 clusters with iron abundances for the majority of elements: Mg, Si, Ca, Mn, and Ni (Table~\ref{tab:apogee_abun}). Aluminum abundances have been derived for 89 systems (the missing cluster, IC~1805, has only one star). Several stars have been rejected in the determination of chromium content so that the abundances of this element have been determined for 84 systems. Sodium is the element for which we reject more stars, with Na abundances obtained for only 65 of the 90 clusters. The abundance of Na is determined in APOGEE from two weak and probably blended lines, that are easily measured only in GK giants \citep[see][for details]{apogeeabundances}. Therefore, Na abundances cannot be determined for many stars in the sample because they are outside this range.

As before, the values obtained here should be used with caution. Only the abundances obtained for at least 4 stars and with small internal dispersion can be considered reliable. Owing to the large heterogeneity of the abundance determinations available in the literature we have not tried a comparison. 

\begin{figure*}
\centering
\includegraphics[scale=0.8]{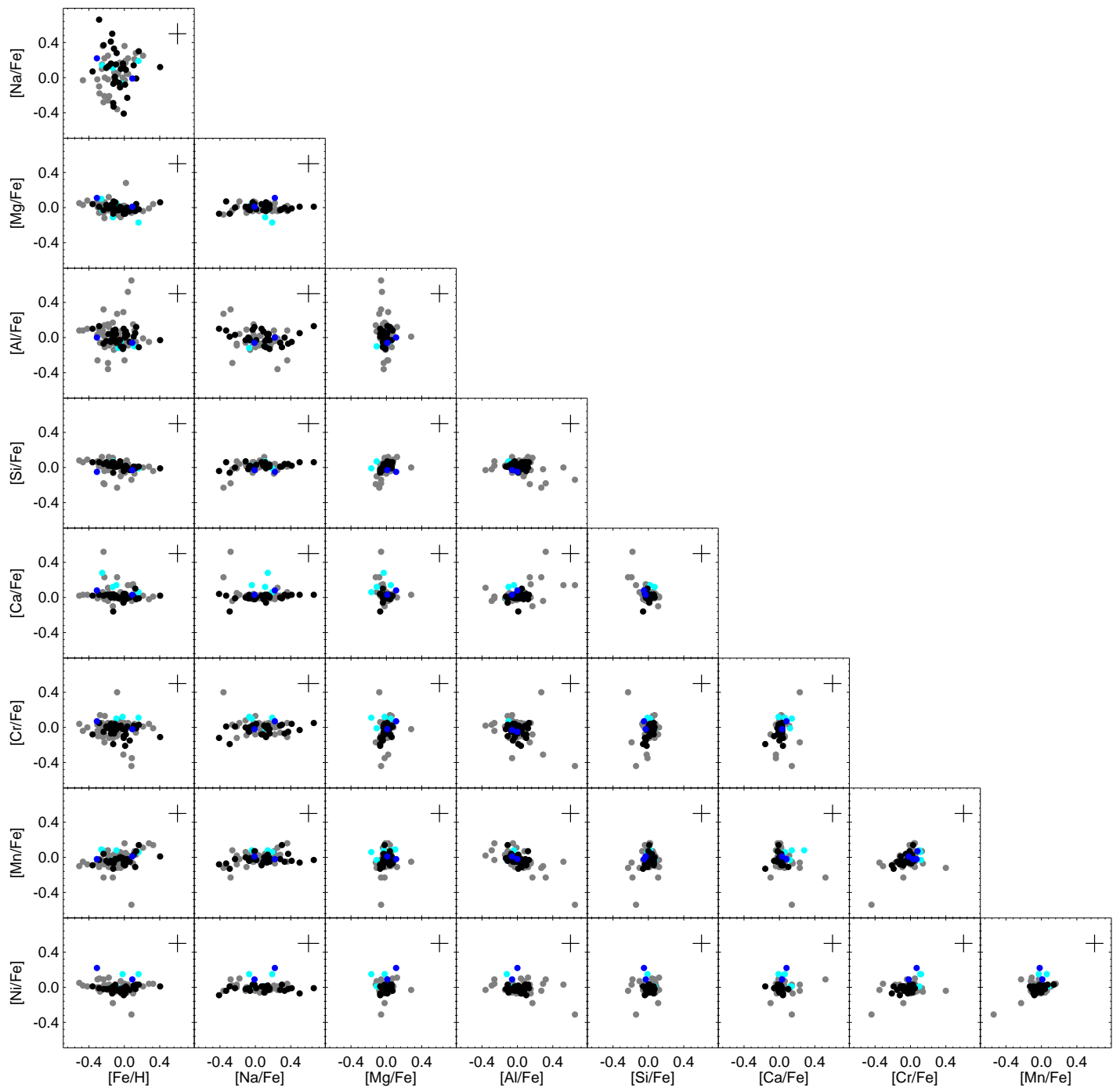}
\caption{Plot of the abundance ratios for APOGEE and GALAH clusters. Grey and black circles are APOGEE clusters with less and more than 4 stars in the determination of [Fe/H], respectively. Cyan and blue circles are GALAH clusters with less and more than 4 stars in the determination of [Fe/H], respectively. Typical error bars have been plotted in top-right corners.}
\label{fig:apogee_abun}
\end{figure*}

In Fig.~\ref{fig:apogee_abun} we show trends of the abundance ratios obtained with [Fe/H] and among all other elements. The most discrepant values are due to clusters with less than 4 stars sampled (grey points). Clusters with more than 4 stars analysed show, in general, a scatter compatible with the typical uncertainties. The exceptions are Na and in less degree Cr. The well know differences of Na abundances between dwarfs and giants due to extra-mixing can explain this scatter. There is one cluster, NGC~2168, for which we obtained [Ca/Fe]=-0.17\,dex from 10 stars. Although this value is lower than the bulk, its dispersion of $\sigma_{[Ca/Fe]}$=0.21\,dex (probably due to the difficulties in measuring main-sequence stars in APOGEE) makes it still compatible with the majority.

\subsection{GALAH}

The GALAH DR2 provides abundances for 23 chemical elements including iron \citep[see][]{buder18}. For homogeneity with the APOGEE sample, we computed average abundances for the same 8 elements than above. Following the \citet{buder18} recommendations we only use those stars that do not have problems in the determination of the abundance of a given element (i.e. {\em flag\_X\_FE=0}). Moreover, we also excluded those clusters with $v\sin i>$20\,km\,s$^{-1}$ because the values obtained for these stars are not reliable. As a result of this, abundances have been determined for only 7 clusters and not for all elements. The obtained values are listed in Table~\ref{A3}.

The GALAH ratios have been plotted in Fig.~\ref{fig:apogee_abun} as cyan (less than 4 stars) and blue (more than 4 objects) filled circles. Although the GALAH sample is small, we see that there are no significant differences between APOGEE and GALAH samples for any of the elements studied. Only the metal-poor cluster NGC~2243 seems to have a large [Ni/Fe] ratio, +0.22\,dex ($\sigma_{[Ni/Fe]}$=0.14\,dex) obtained from 5 stars, in comparison with other APOGEE clusters of the same [Fe/H]. This cluster is the only one in common with the APOGEE sample with reliable abundances in GALAH. For direct comparison, in the case of APOGEE we obtained [Ni/Fe]=0.01\,dex from a single star with e$_{[Ni/Fe]}$=0.02\,dex. Also for Si we found a difference larger than the the uncertainties: [Si/Fe]=0.09 and -0.04\,dex from APOGEE and GALAH, respectively. Again the APOGEE value has been derived from a single star with e$_{[Si/Fe]}$=0.03\,dex while the GALAH one has been determined from 4 stars with $\sigma_{[Ni/Fe]}$=0.04\,dex. For the other elements studied the ratios obtained from APOGEE and GALAH samples are in agreement within the uncertainties. 

\section{Galactic trends}\label{sect:trends}

\begin{figure}
\centering
\includegraphics[scale=0.35]{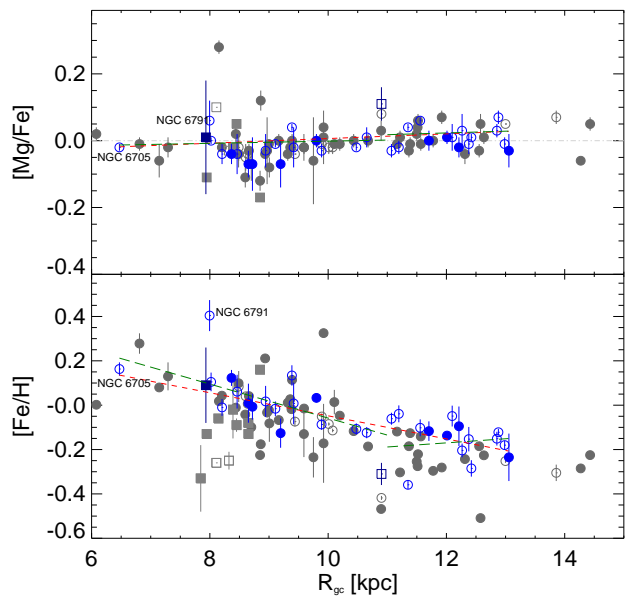}
\caption{Gradient in [Mg/Fe] (top) and [Fe/H] (bottom) as a function of R$_{gc}$ for APOGEE (circles) and GALAH (triangles) clusters. Grey symbols are clusters with less than four stars. Blue symbols are clusters with 4 or more stars sampled. Filled symbols are the clusters for which this is the first metallicity determination from high resolution spectroscopy. Green and orange lines show different linear fits (see text for details). The positions of NGC~6705 and NGC~6791 have been marked.}
\label{fig:trend_rgc}
\end{figure}

As commented in Sect.~\ref{sect:intro}, information about the chemical composition of OCs is necessary to address a variety of astrophysical topics. A clear example of the applicability of the sample obtained in this work is the study of the chemical gradients in the Galactic disk. Generally the chemical gradients in the Galactic disk are studied using iron \citep[e.g.][]{netopil16}, which is produced in approximately equal measure by core collapse and type Ia supernovae. Additionally, we present here the behaviour of magnesium as best representative of the $\alpha$-elements. Not only the production of magnesium is dominated by core collapse supernovae, but also the Mg abundances derived by APOGEE and GALAH show the best agreement with external measurements in comparison with other elements of this group \citep{apogeeabundances,buder18}. Given the much larger number of clusters involved, the APOGEE data dominate the following discussion. The run of [Fe/H] as a function of Galactocentric distance, R$_{gc}$, is plotted in the bottom panel of Fig.~\ref{fig:trend_rgc}, while the top panel shows the behaviour of [Mg/Fe] with R$_{gc}$. Galactocentric distances have been computed by \citet{tristan18b} from \textit{Gaia}-DR2 parallaxes; we refer the reader to that paper for details about the distance determination.

The clusters with 4 or more stars, with more trustful measurements (blue symbols), cover a range in R$_{gc}$ between $\sim$6.5 and $\sim$13\,kpc. Grey symbols are clusters with less reliable measurements. There is no full agreement about the slope of the gradient in this R$_{gc}$ range and we can found in the literature values  between $\frac{d[FeH]}{dR_{gc}}\sim$-0.035\,dex\,kpc$^{-1}$ \citep{cunha2016} and -0.1\,dex\,kpc$^{-1}$ \citep{jacobson16}. Using all the clusters in the APOGEE and GALAH samples with at least 4 stars we found $\frac{d[Fe/H]}{dR_{gc}}$=-0.052$\pm$0.003\,dex\,kpc$^{-1}$ (red dashed line). The slope of the gradients may depend on the presence of the innermost cluster in the sample, NGC~6705, and the most metal-rich one, NGC~6791. Both are peculiar clusters. NGC~6705 is a young metal-rich system \citep[see e.g.][]{tristan14} with an unexpected high $\alpha$-elements abundance \citep{occasoIII,magrini17}. On the contrary, NGC~6791 is an intriguing old very metal-rich and massive system located almost 1\,kpc above the Galactic plane. It has been suggested that NGC~6791 has likely migrated to its current location from its birth position \citep{linden17} or even has an extragalactic origin \citep{carraro2006} although both claims are disputed. If we exclude these two clusters from the analysis, the [Fe/H] gradient flattens to -0.047$\pm$0.004\,dex\,kpc$^{-1}$. This does not imply that former value is preferred. It only shows how the gradient changes as a function of the outliers. 
 %This flattening can be due to the fact suggested by previous studies that the gradient flattens for R$_{gc}\geq$11\,kpc \citep[e.g][]{carrera&pancino2011,netopli16}. In fact our sample contains a significant number of systems at these distances. 

If we separate the clusters in two groups inside and outside R$_{gc}=$11\,kpc we find $\frac{d[Fe/H]}{dR_{gc}}$=-0.077$\pm$0.007\,dex\,kpc$^{-1}$ in the inner region and $\frac{d[Fe/H]}{dR_{gc}}$=0.018$\pm$0.009\,dex\,kpc$^{-1}$ for the outer region. A similar result has been reported previously \citep[e.g.][among many]{carrera&pancino2011,andreuzzi2011,occam1,cantat_gaudin2016}.
%with a median [Fe/H] of -0.14\,dex with a standard deviation of 0.09\,dex. 
If we exclude the two metal-poor cluster at R$_{gc}\sim$11\,kpc (NGC~2243 and Trumpler~5) the slope increases until $\frac{d[Fe/H]}{dR_{gc}}$=-0.04$\pm$0.01\,dex\,kpc$^{-1}$. Therefore, the behaviour in the outermost region is highly dependent on these clusters. All these results are in good agreement with \citet{occam2} who also used chemical abundances obtained from APOGEE DR14 with a different cluster membership selection. Furthermore, the observed gradient may change as a function of the age of the clusters used in the analysis \citep[e.g.][]{friel02,andreuzzi2011,carrera&pancino2011,jacobson16}. However, age is yet unknown for a large fraction of our clusters and we postpone to a forthcoming paper a detailed analysis of the evolution of the gradient with the age of the clusters.

The run of $\alpha$-element abundances as a function of Galactocentric radius is still an open discussion \citep[e.g.][]{donati15,cantat_gaudin2016}. \citet{occam2} reported a mild positive gradient for magnesium and other $\alpha$-elements such as oxygen or silicon. On contrast, \citet{yong12} and \citet{friel2014} did not found any dependence with R$_{gc}$. A similar result has been found by \citet{cantat_gaudin2016} and also by \citet{magrini17} using OCs and field stars homogeneously analysed by the $Gaia$-ESO survey. The top panel of Fig.~\ref{fig:trend_rgc} does not show a clear trend. The slope of the linear fit (red dashed line) to the whole range of R$_{gc}$ is $\frac{d[Mg/Fe]}{dR_{gc}}$=0.003$\pm$0.002\,dex\,kpc$^{-1}$. The clusters within R$_{gc}\sim$10\,kpc have, in general, a [Mg/Fe] below the solar one.  This includes the innermost cluster, NGC~6705, for which \citet{occasoIII} has reported a higher value of [Mg/Fe]=+0.14$\pm$0.07\,dex, in agreement with the $Gaia$-ESO result of $+0.10\pm0.07$ \citep{magrini17}. There is a group of clusters at $\sim$8\,kpc which [Mg/Fe] are compatible with the solar one within the uncertainties. At R$_{gc}\sim$8.5\,kpc the [Mg/Fe] is clearly lower than the solar. From there the [Mg/Fe] ratio increases until $\sim$10\,kpc and it flattens from thereof. This behaviour has been reported previously in the literature \citep[e.g.][]{cantat_gaudin2016,magrini17} and it has been predicted by Galactic chemical evolution models \citep[e.g.][]{minchev2014,kubryk2015a,kubryk2015b,grisoni2018}.

\begin{figure}
\centering
\includegraphics[scale=0.35]{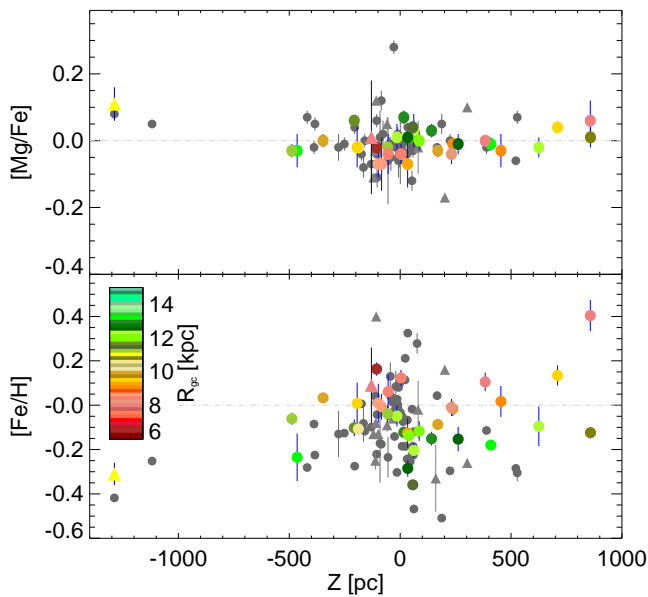}
\caption{The run of [Mg/Fe] (top) and [Fe/H] (bottom) as a function of the distance to the Galactic plane  $Z$. Symbols  are as in Fig.~\ref{fig:trend_rgc}.}
\label{fig:trend_z}
\end{figure}

The existence of a vertical gradient is also controversial. Several authors do not find any trend of [Fe/H] with the distance to the Galactic plane, \textit{Z} \citep[e.g.][]{jacobson2011,carrera&pancino2011}. Instead, other studies reported the existence of a vertical gradient with a slope between -0.34 and -0.25\,dex\,kpc$^{-1}$ \citep[e.g][]{piatti1995,carraro1998}. In Fig.\ref{fig:trend_z} we plot [Fe/H] and [Mg/Fe] as a function of \textit{Z} (bottom and upper panel, respectively). In both cases there are no hints of the existence of vertical gradients in agreement with most previous studies. We also confirm that clusters located at larger Galactocentric distances cover a larger range of \textit{Z}. The exception is NGC~6791 located at R$_{gc}\sim$8\,kpc and with a height above the plane of $Z\sim$900\,pc. We have already commented that this is a peculiar and not well understood system.

\section{Summary}\label{sect:conclusion}

Using \textit{Gaia}~DR2 \citet{tristan18b} determined astrometric membership probabilities for stars belonging to 1229 open clusters. We have cross-matched this catalogue with the latest data releases of two of the largest Galactic high-resolution spectroscopic surveys, APOGEE and GALAH, with the goal of finding high probability OC members.

\begin{itemize}
\item In the case of APOGEE we have detected stars belonging to 131 clusters for which we have determined average radial velocities. For 46 systems our determination is based on 4 or more stars. To our knowledge this is the first radial velocity determination for 16 systems. For the other clusters there is a good agreement between the obtained values and those available in the literature. [Fe/H] has been obtained for 90 open clusters, almost two thirds of them without previous determinations in the literature. Finally, for the same 90 clusters we have also determined abundances for six elements: Na, Mg, Al, Si, Ca, Cr, Mn, and Ni.
\item In the case of GALAH we have found stars belonging to 14 clusters for which we have determined both radial velocities and iron abundances. Except for two clusters, NGC~2243 and NGC~2548, the GALAH sample is composed by main-sequence stars, in some cases with significant $v\sin i$ values. These 14 clusters have previous radial velocity determination from \textit{Gaia}~DR2. Excluding the two clusters in common with APOGEE sample, nine of these systems do not have previous determinations in the literature from high resolution spectroscopy. For seven clusters we have determined abundances for the same chemical elements that in the case of APOGEE.

\end{itemize}

In summary, to our knowledge this is the first RV determination from high resolution spectra for 16 open clusters. In the same way, we provide for the first time iron abundances for 39 open clusters, 30 from APOGEE and 9 from GALAH, respectively.

The obtained catalogue of chemical abundances has been used to investigate the existence of radial and vertical trends using the distances computed from \textit{Gaia}~DR2. Our findings are in agreement with previous investigations where the radial [Fe/H] gradient appears to flatten in the outer region. We do not find any hint, at least above the 1$\sigma$ level, of the existence of a vertical metallicity gradient. 

\begin{acknowledgements}
This work has made use of data from the European Space Agency (ESA) mission
{\it Gaia} (\url{https://www.cosmos.esa.int/gaia}), processed by the {\it Gaia}
Data Processing and Analysis Consortium (DPAC,
\url{https://www.cosmos.esa.int/web/gaia/dpac/consortium}). Funding for the DPAC
has been provided by national institutions, in particular the institutions
participating in the {\it Gaia} Multilateral Agreement.
This work has made use of APOGEE data.
Funding for the Sloan Digital Sky Survey IV has been provided by the Alfred P. Sloan Foundation, the U.S. Department of Energy Office of Science, and the Participating Institutions. SDSS-IV acknowledges
support and resources from the Center for High-Performance Computing at
the University of Utah. The SDSS web site is www.sdss.org.

SDSS-IV is managed by the Astrophysical Research Consortium for the 
Participating Institutions of the SDSS Collaboration including the 
Brazilian Participation Group, the Carnegie Institution for Science, 
Carnegie Mellon University, the Chilean Participation Group, the French Participation Group, Harvard-Smithsonian Center for Astrophysics, 
Instituto de Astrof\'isica de Canarias, The Johns Hopkins University, 
Kavli Institute for the Physics and Mathematics of the Universe (IPMU) / 
University of Tokyo, the Korean Participation Group, Lawrence Berkeley National Laboratory, 
Leibniz Institut f\"ur Astrophysik Potsdam (AIP),  
Max-Planck-Institut f\"ur Astronomie (MPIA Heidelberg), 
Max-Planck-Institut f\"ur Astrophysik (MPA Garching), 
Max-Planck-Institut f\"ur Extraterrestrische Physik (MPE), 
National Astronomical Observatories of China, New Mexico State University, 
New York University, University of Notre Dame, 
Observat\'ario Nacional / MCTI, The Ohio State University, 
Pennsylvania State University, Shanghai Astronomical Observatory, 
United Kingdom Participation Group,
Universidad Nacional Aut\'onoma de M\'exico, University of Arizona, 
University of Colorado Boulder, University of Oxford, University of Portsmouth, 
University of Utah, University of Virginia, University of Washington, University of Wisconsin, 
Vanderbilt University, and Yale University.
This work has made use of GALAH data,
based on data acquired through the Australian Astronomical Observatory, under programmes: A/2013B/13 (The GALAH pilot survey); A/2014A/25, A/2015A/19, and A2017A/18 (The GALAH survey). We acknowledge the traditional owners of the land on which the AAT stands, the Gamilaraay people, and pay our respects to elders past and present.This research has made use of Vizier and SIMBAD, operated at CDS, Strasbourg,
France,  NASA's Astrophysical Data System, and {\sc TOPCAT} (http://www.starlink.ac.uk/topcat/, \citealt{topcat}).  This research made use of the cross-match service provided by CDS, Strasbourg. This work was partly supported by the
MINECO (Spanish Ministry of Economy) through grant ESP2016-80079-C2-1-R
(MINECO/FEDER, UE) and MDM-2014-0369 of ICCUB (Unidad de Excelencia
'Mar\'{\i}a de Maeztu'). 
A.B. acknowledges funding from PREMIALE 2015 MITiC.
\end{acknowledgements}

%-------------------------------------------------------------------
\bibliographystyle{aa} % style aa.bst
\bibliography{biblio_apogee_galah.bib}

\appendix
\section{Tables}

{\fontsize{6}{2}\selectfont
\begin{landscape}
\onecolumn \centering
\setlength{\tabcolsep}{0.5mm}
\begin{longtable}{lccccccccccccccccccccc}
\caption{Information on APOGEE DR14 stars in common with candidate members in \textit{Gaia}~DR2 \citep{tristan18b}.} \label{A1} \\
\hline
Cluster & star\_id & source\_id & RA & DEC & $\varpi$ & $\sigma_{\varpi}$ & $\mu_{\alpha*}$ & $\sigma_{\mu_{\alpha*}}$ & $\mu_{\delta}$ & $\sigma_{\mu_{\delta}}$ & $G$ & $G_{BP}-G_{RP}$ & \textit{p} & RV & $\sigma_{RV}$ & Teff & $\sigma_{\rm Teff}$ & $\log g$ & $\sigma_{\log g}$ & [Fe/H] & $\sigma_{[Fe/H]}$ \\
\hline
\endfirsthead
  ASCC~124 & 2M22450300+4613588 & 1982689777242358912 & 341.26249155537 & 46.23300470968 & 1.3276718736945818 & 0.04611155225642544 & 0.2996917390783982 & 0.06885027846654153 & -2.0589048051205565 & 0.0662293265210654 & 9.92051 & 0.12712765 & & 0.5 & -18.1821 & 0.673 & & & & &  \\
\hline
\end{longtable}
\twocolumn
\end{landscape}
}

{\fontsize{6}{2}\selectfont
\begin{landscape}
\onecolumn \centering
\setlength{\tabcolsep}{0.5mm}
\begin{longtable}{lcccccccccccccccccccccccc}
\caption{Information on GALAH DR2 stars in common with candidate members in \textit{Gaia}~DR2 \citep{tristan18b}.} \label{A2} \\
\hline
CLUSTER & star\_id & source\_id & RA & DEC & $\varpi$ & $\sigma_{\varpi}$ & $\mu_{\alpha*}$ & $\sigma_{\mu_{\alpha*}}$ & $\mu_{\delta}$ & $\sigma_{\mu_{\delta}}$ & $G$ & $G_{BP}-G_{RP}$ & \textit{p} & RV & $\sigma_{RV}$ & f\_c & Teff & $\sigma_{\rm Teff}$ & $\log g$ & $\sigma_{\log g}$ & [Fe/H] & $\sigma_{[Fe/H]}$ & $v\sin i$ & $\sigma_{v\sin i}$ \\
\hline
\endfirsthead
ASCC~16 & 05251613+0231412 & 3234352820498417024 & 81.3172531 & 2.5281074 & 2.809 & 0.058 & 1.187 & 0.09 & -0.546 & 0.075 & 10.041 & 0.231 & 0.6 & 19.85 & 0.18 & 1 & 7557 & 48 & 4.45 & 0.14 & -0.25 & 0.06 & 23.92 & 0.83\\

\hline
\end{longtable}
\twocolumn
\end{landscape}
}

\begin{landscape}
\onecolumn \centering
\setlength{\tabcolsep}{0.5mm}
\begin{table*}
\centering
\caption{Individual stars in the GALAH DR2 clusters retained in the analysis.}
\begin{tabular}{lcccccccccc}
\hline
Cluster & star\_id & source\_id & $G$ & $G_{BP}-G_{RP}$ & RV & T$_{\rm eff}$ & $\log g$ & [Fe/H] & $v\sin i$ & flag$_{\rm cannon}$ \\
\hline
  ASCC~16 & 05242648+0126492 & 3222160885813451776 & 11.044314 & 0.81302357 & 22.0104529617 & 5870.960626083559 & 4.205298815473945 & -0.07787835749222094 & 33.6240327478691 & 1\\

\hline
\end{tabular}
\label{selected_G}
\end{table*}
\twocolumn
\end{landscape}

{\fontsize{7}{1}\selectfont
\begin{landscape}
\onecolumn \centering
\setlength{\tabcolsep}{0.5mm}
\begin{longtable}{lcccccccccccccccccccccccccccccccc}
\caption{Abundances obtained for OCs in the APOGEE sample.}\label{tab:apogee_abun}\\
\hline
Cluster & Na & $\sigma_{Na} $ & $e_{Na}$ & Nr$_{Na}$ & Mg & $\sigma_{Mg}$ & $e_{Mg}$ &
Nr$_{Mg}$ & Al & $\sigma_{Al}$ & $e_{Al}$ & Nr$_{Al}$ & Si] & $\sigma_{Si}$  & e$_{Si}$ & Nr$_{Si}$ & Ca & $\sigma_{Ca}$ & e$_{Ca}$ & Nr$_{Ca}$ & Cr & $\sigma_{Cr}$ & e$_{Cr}$ & Nr$_{Cr}$ & Mn & $\sigma_{Mn}$ & e$_{Mn}$ & Nr$_{Mn}$ & Ni & $\sigma_{Ni}$ & e$_{N}$ & Nr$_{Ni}$\\
\hline
\endfirsthead
  ASCC~21 & & & & & -0.07 & 0.04 & 0.01 & 10 & 0.04 & 0.12 & 0.04 & 10 & -0.02 & 0.06 & 0.02 & 10 & 0.04 & 0.07 & 0.03 & 8 & -0.21 & 0.12 & 0.08 & 2 & -0.09 & 0.04 & 0.01 & 10 & -0.07 & 0.05 & 0.02 & 10\\

\hline
\end{longtable}
\twocolumn
\end{landscape}
}

{\fontsize{7}{1}\selectfont
\begin{landscape}
\onecolumn \centering
\setlength{\tabcolsep}{0.5mm}
\begin{longtable}{lcccccccccccccccccccccccccccccccc}
\caption{Mean [X/Fe] abundance ratios for the GALAH clusters.} \label{A3} \\
\hline
Cluster & Na & $\sigma_{Na} $ & $e_{Na}$ & Nr$_{Na}$ & Mg & $\sigma_{Mg}$ & $e_{Mg}$ &
Nr$_{Mg}$ & Al & $\sigma_{Al}$ & $e_{Al}$ & Nr$_{Al}$ & Si] & $\sigma_{Si}$  & e$_{Si}$ & Nr$_{Si}$ & Ca & $\sigma_{Ca}$ & e$_{Ca}$ & Nr$_{Ca}$ & Cr & $\sigma_{Cr}$ & e$_{Cr}$ & Nr$_{Cr}$ & Mn & $\sigma_{Mn}$ & e$_{Mn}$ & Nr$_{Mn}$ & Ni & $\sigma_{Ni}$ & e$_{N}$ & Nr$_{Ni}$\\
\hline
\endfirsthead
 Alessi~24 & 0.11 & 0.0 & 0.05 & 1 & -0.11 & 0.0 & 0.07 & 1 & -0.1 & 0.0 & 0.04 & 1 & 0.07 & 0.0 & 0.07 & 1 & 0.12 & 0.0 & 0.05 & 1 & -0.01 & 0.0 & 0.05 & 1 & -0.03 & 0.0 & 0.06 & 1 & 0.01 & 0.0 & 0.06 & 1\\

\hline
\end{longtable}
\twocolumn
\end{landscape}
}

{\fontsize{7}{1}\selectfont
\begin{landscape}
\onecolumn \centering
\setlength{\tabcolsep}{0.5mm}
\begin{longtable}{lc|cccc|ccc|ccc|cccc|ccc|c}
\caption{The 131 open clusters in common between APOGEE DR14 and \citet{tristan18b}.\label{tab_apogee_rv_ext}}\\
\hline
Cluster & Star & RV & $\sigma_{RV}$ & e$_{RV}$ & Nr & RV$_{lit}$ & $\sigma_{RV,lit}$ & Nr$_{lit}$ & RV$_{GDR2}$\tablefootmark{b} & $\sigma_{GDR2}$ & Nr$_{GDR2}$ & [Fe/H] & $\sigma_{[Fe/H]}$ & e$_{[Fe/H]}$ & Nr & [Fe/H]$_{lit}$ & $\sigma_{[Fe/H],lit}$ & Nr$_{lit}$ & Ref.\\
 & type\tablefootmark{a}  &\multicolumn{3}{c}{(km~s$^{-1}$)} & & \multicolumn{2}{c}{(km~s$^{-1}$)}&  & \multicolumn{2}{c}{(km~s$^{-1}$)}&  & \multicolumn{3}{c}{(dex)} & & \multicolumn{2}{c}{(dex)}& & \\
\hline
\endfirsthead
  Alessi~20 & MS & -14.89 & & 0.37 & 1 & -11.5 & 0.01 & 2 & -5.04 & 3.3 & 7 &  &  &  &  &  &  &  &  1\\
  ASCC~124 & MS & -23.35 & 0.54 & 0.38 & 2 &  &  &  &  &  &  &  &  &  &  &  &  &  & \\
  ASCC~16 & MS & 17.41 &  & 0.62 & 1 &  &  &  & 23.18 & 3.4 & 15 &  &  &  &  &  &  &  & \\
  ASCC~21 & MS & 16.30 & 5.66 & 1.30 & 19 & 18.57 & 2.12 & 9 & 18.7 & 4.45 & 9 & 0.01 & 0.09 & 0.03 & 10 & & & & 2\\
  Basel~11b & RGB & 2.68 & 0.06 & 0.04 & 2 &  &  &  & 3.43 & 1.46 & 3 & 0.014 & 0.05 & 0.04 & 2 &  &  &  &  \\
  Berkeley~17 & RGB & -73.34 & 0.41 & 0.13 & 9 & -73.4 & 0.4 & 7 & -71.95 & 1.77 & 7 & -0.10 & 0.04 & 0.01 & 9 & -0.11 & 0.03 & 7 & 3\\
  Berkeley~19 & RGB & 17.44 & 0.0 & 0.08 & 1 &  &  &  & 17.65 & 0.42 & 1 & -0.22 & & 0.01 & 1 &  &  &  & \\
  Berkeley~29 & RGB & 25.27 & 0.53 & 0.30 & 3 & 24.8 & 1.13 & 11 & 50.58 & 0.94 & 1 &  &  &  &  &  &  &  & 1\\
  Berkeley~31 & RGB & 57.87 & 0.65 & 0.46 & 2 & 61.0 & 3.75 & 17 &  &  &  & -0.305 & 0.037 & 0.026 & 2 & -0.31 & 0.06 & 2 & 1,12\\
  Berkeley~33 & RGB & 77.80 & 0.56 & 0.28 & 4 & 76.6 & 0.5 & 5 & 78.82 & 1.12 & 4 & -0.23 & 0.11 & 0.05 & 4 &  & &  & 2\\
  Berkeley~43 & RGB & 29.712 &  & 0.136 & 1 &  &  &  & 29.15 & 1.42 & 9 & 0.00 & & 0.01 & 1 &  & &  & \\
  Berkeley~53 & RGB & -35.77 & 0.82 & 0.29 & 8 & -36.3 & 0.5 & 4 & -34.9 & 1.81 & 7 & -0.02 & 0.03 & 0.01 & 6 & 0.00 & 0.02 & 5 & 3\\
  Berkeley~66 & RGB & -50.14 & 0.21 & 0.09 & 5 & -50.1 & 0.3 & 6 & -76.19 & 3.28 & 1 & -0.12 & 0.01 & 0.01 & 5 & -0.13 & 0.02 & 6 & 3\\
  Berkeley~71 & both & -9.63 & 0.38 & 0.16 & 6 & -8.7 & 2.3 & 7 & -26.7 & 4.19 & 1 & -0.20 & 0.03 & 0.01 & 6 & -0.2 & 0.03 & 7 & 3\\
  Berkeley~7 & MS & 18.73 & & 1.15 & 1 &  &  &  &  &  &  &  &  &  &  &  &  &  & \\
  Berkeley~85 & both & -34.93 & 0.37 & 0.12 & 9 &  &  &  & -33.72 & 1.4 & 18 &  &  &  &  &  &  &  & \\
  Berkeley~86 & MS & 151.90 &  & 0.83 & 1 & -25.54 & 2.6 & 2 &  &  &  &  &  &  &  &  &  &  & 2\\
  Berkeley~87 & MS & 1.98 & 5.42 & 2.049 & 7 & -8.6 & 1.88 & 2 & -7.5 & 0.0 & 1 &  &  &  &  &  &  &  & 1\\
  Berkeley~91 & RGB & -51.37 & & 0.07 & 1 &  &  &  & -45.11 & 2.41 & 2 & 0.11 & & 0.01 & 1 &  & & &\\
  Berkeley~98 & RGB & -67.22 & 3.27 & 1.46 & 5 &  &  &  &  &  &  & 0.033 & 0.02 & 0.01 & 5 &  & & & \\
  Berkeley~9 & both & -18.47 & 1.77 & 0.79 & 5 &  &  &  & -17.4 & 0.51 & 3 & -0.172 & 0.18 & 0.10 & 3 &  & &  &  \\
  Collinder~359 & MS & 3.61 & 0.05 & 0.03 & 3 & -5.37 & 2.46 & 10 & 5.28 & 3.25 & 12 &  &  &  &  &  &  &  & 2\\
  Collinder~69 & MS & 27.45 & 4.00 & 0.37 & 118 & 31.38 & 1.42 & 4 & 29.1 & 3.24 & 29 & -0.01 & 0.06 & 0.01 & 24 &  & & & 2 \\
  Collinder~95 & MS & 16.24 & 0.43 & 0.31 & 2 & -11.0 & 7.4 & 1 & 8.47 & 4.74 & 2 & -0.03 & 0.02 & 0.01 & 2 &  & & & 2\\
  Czernik~18 & MS & -15.01 & 0.49 & 0.28 & 3 &  &  &  &  &  &  & -0.13 & 0.10 & 0.06 & 3 &  &  &  & \\
  Czernik~20 & RGB & 30.85 & 0.75 & 0.28 & 7 & -29.97 & 10.0 & 1 &  &  &  & -0.12 & 0.04 & 0.02 & 7 &  &  &  & 2\\
  Czernik~21 & RGB & 44.92 & 0.67 & 0.39 & 3 &  &  &  & 45.52 & 1.64 & 2 & -0.24 & 0.01 & 0.01 & 3 &  & & & \\
  Czernik~23 & RGB & 17.75 & & 0.04 & 1 &  &  &  & 13.09 & 10.31 & 2 & -0.25 & & 0.01 & 1 &  & & & \\
  Czernik~30 & RGB & 81.29 & 0.50 & 0.35 & 2 & 79.9 & 1.5 & 17 &  &  &  & -0.285 & 0.02 & 0.01 & 2 & & & & 2\\
  Dolidze~3 & MS & -7.65 & & 1.87 & 1 &  &  &  &  &  &  &  &  &  &  &  &  &  &  \\
  Dolidze~5 & MS & -37.78 & & 0.03 & 1 & -35.0 & 0.0 & 1 & -20.4 & 0.33 & 4 &  &  &  &  &  &  &  & 2 \\
  Feibelman~1 & MS & -6.18 & & 2.50 & 1 & -8.9 & 0.74 & 1 &  &  &  &  &  &  &  &  &  &  & 2 \\
  FSR~0306 & RGB & -38.41 & & 0.10 & 1 &  &  &  & -34.63 & 4.13 & 3 & 0.21 & & 0.01 & 1 &  & & & \\
  FSR~0336 & MS & -26.58 & & 3.12 & 1 &  &  &  & -27.32 & 1.88 & 2 &  &  &  &  &  &  &  & \\
 FSR~0496 & RGB & -23.26 & & 0.16 & 1 &  &  &  & -22.98 & 1.11 & 12 & -0.07 &  & 0.01 & 1 &  & & & \\
  FSR~0542 & RGB & -74.09 & & 0.16 & 1 &  &  &  & -63.93 & 0.0 & 1 & -0.19 & & 0.01 & 1 &  & & &  \\
  FSR~0667 & RGB & 2.11 & 0.07 & 0.05 & 2 &  &  &  & 2.5 & 3.55 & 2 & 0.03 & 0.01 & 0.01 & 2 &  & & & \\
  FSR~0716 & RGB & 7.33 &  & 0.05 & 1 &  &  &  & 11.18 & 0.0 & 1 & -0.30 & & 0.01 & 1 &  &  &  & \\
  FSR~0826 & RGB & -4.08 & & 0.07 & 1 &  &  &  &  &  &  & -0.14 &  & 0.01 & 1 &  &  &  & \\
  FSR~0883 & RGB & 15.07 & & 0.07 & 1 & -5.0 & 20.5 & 3 & 15.28 & 1.11 & 2 & -0.18 & & 0.01 & 1 &  & & & 2 \\
  FSR~0941 & RGB & 20.17 & & 0.05 & 1 &  &  &  &  &  &  & -0.23 & & 0.01 & 1 &  & &  & \\
  FSR~0942 & RGB & 29.06 & & 0.06 & 1 &  &  &  & 31.88 & 0.0 & 1 & -0.27 & & 0.01 & 1 &  & & & \\
  FSR~1063 & RGB & 56.81 & & 0.13 & 1 &  &  &  & 59.76 & 1.2 & 1 & -0.12 & & 0.01 & 1 &  &  &  &  \\
  Gulliver~23 & RGB & -6.21 & & 0.09 & 1 &  &  &  & -3.33 & 1.47 & 9 &  &  &  &  &  &  &  &  \\
  Gulliver~6 & MS & 27.26 & & 0.01 & 1 &  &  &  & 28.35 & 5.36 & 15 & -0.10 & & 0.01 & 1 &  & &  & \\
  Haffner~4 & RGB & 59.65 & & 0.16 & 1 &  &  &  & 57.05 & 0.0 & 1 & -0.126 & & 0.008 & 1 &  & &  & \\
  IC~1369 & RGB & -48.83 & 0.09 & 0.05 & 3 &  &  &  & -48.25 & 2.17 & 9 & -0.02 & 0.01 & 0.01 & 3 &  &  &  & \\
  IC~166 & RGB & -39.85 & 2.07 & 0.48 & 19 & -40.5 & 1.5 & 15 & -39.82 & 3.81 & 4 & -0.05 & 0.05 & 0.01 & 19 & -0.08 & 0.05 & 13 & 2,8\\
  IC~1805 & MS & -40.60 & & 0.21 & 1 & -48.5 & 6.11 & 6 & -42.39 & 1.28 & 2 & 0.32 & & 0.01 & 1 &  &  & & 1 \\
  IC~348 & MS & 15.83 & 2.61 & 0.25 & 11 & 18.1 & 3.5 & 4 & 19.71 & 1.57 & 6 &  &  &  &  &  &  &  & 2 \\
  IC~4996 & MS & 78.51 & & 0.3 & 1 & -2.5 & 5.75 & 4 & -29.11 & 2.94 & 1 &  &  &  &  &  &  &  & 2\\
  IC~5146 & MS & -5.48 & 0.35 & 0.25 & 2 & -3.6 & 1.65 & 2 & -2.12 & 5.64 & 2 & -0.01 & 0.01 & 0.01 & 2 &  & & & 1\\
  King~15 & RGB & -64.80 & & 0.08 & 1 &  &  &  & -65.81 & 1.27 & 3 & -0.05 & & 0.01 & 1 & & &  & \\
  King~1 & RGB & -52.43 & & 0.01 & 1 & -53.1 & 3.1 & 28 & -53.2 & 2.0 & 27 & -0.02 &  & 0.01 & 1 &  &  & & 5\\
  King~2 & RGB & -135.79 & & 0.04 & 1 & -144.25 & 5.92 & 7 &  &  &  & -0.28 & & 0.01 & 1 &  & & & 2 \\
  King~5 & RGB & -43.78 & 0.93 & 0.42 & 5 & -44.3 & 1.5 & 5 & -44.39 & 2.39 & 14 & -0.11 & 0.02 & 0.01 & 5 & -0.11 & 0.02 & 5 & 3\\
  King~7 & both & -10.29 & 1.43 & 0.41 & 12 & -11.9 & 2.0 & 4 & -9.57 & 1.25 & 12 & -0.04 & 0.04 & 0.01 & 7 & -0.05 & 0.02 & 4 & 3\\
  King~8 & RGB & -1.18 & & 0.12 & 1 & -1.0 & 3.0 & 2 & 22.8 & 0.0 & 1 &  &  &  &  &  &  &  & 2\\
  Koposov~36 & MS & 4.23 & 9.33 & 4.17 & 5 &  &  &  & 52.26 & & 1 &  &  &  &  &  &  &  & \\
  Koposov~63 & RGB & -7.09 & & 0.11 & 1 &  &  &  & 13.68 & & 1 & -0.51 & & 0.01 & 1 &  & &  & \\
  Kronberger~57 & RGB & -1.57 & & 0.05 & 1 &  &  &  &  &  &  & 0.02 & & 0.01 & 1 &  & &  & \\
  L~1641S & MS & 22.11 & 2.08 & 0.32 & 43 &  &  &  &  &  &  &  &  &  &  &  &  &  & \\
  Melotte~20 & MS & 0.21 & 2.04 & 0.59 & 12 & -1.4 & 0.65 & 90 & -0.1 & 2.05 & 126 & 0.08 & 0.07 & 0.04 & 3 & 0.03 & 0.11 & 1 & 1,9\\
  Melotte~22 & MS & 6.57 & 3.06 & 0.17 & 308 & 5.5 & 0.33 & 106 & 5.92 & 1.35 & 212 & 0.061 & 0.08 & 0.01 & 217 & -0.01 & 0.05 & 12 & 1,10\\
  Melotte~71 & RGB & 50.83 & 0.42 & 0.19 & 5 & 50.1 & 0.14 & 11 & 51.26 & 1.65 & 21 & -0.09 & 0.02 & 0.01 & 5 & -0.09 & 0.17 & 1 & 1,9\\
  Negueruela~1 & MS & -71.39 & & 0.4 & 51 & -60.3 & 0.01 & 2 & -5.02 & 2. 94 & 1 &  &  &  &  &  &  &  & 2 \\
  NGC~1193 & RGB & -85.16 & 0.18 & 0.13 & 2 & -82.0 & 10.39 & 4 & -83.24 & 0.51 & 1 & -0.25& 0.01 & 0.01 & 2 & -0.22 & 0.01 & 2 & 10\\
  NGC~1245 & RGB & -29.38 & 0.47 & 0.09 & 27 & -29.2 & 0.8 & 23 & -29.6 & 1.06 & 19 & -0.06 & 0.903 & 0.01 & 27 & -0.06 & 0.02 & 23 & 3\\
  NGC~129 & MS & -35.67 & & 0.09 & 1 & -38.0 & 0.82 & 5 & -37.89 & 6.84 & 2 &  &  &  &  &  &  &  & 1\\
  NGC~1333 & MS & 15.76 & 2.67 & 0.42 & 40 &  &  &  & 1.92 & 5.91 & 1 & -0.04 &  & 0.01 & 1 &  &  & \\
  NGC~136 & RGB & -52.13 & & 0.06 & 1 &  &  &  & -47.38 & 5.46 & 1 & -0.22 & & 0.01 & 1 &  &  &  & \\
  NGC~1579 & MS & 11.02 & & 1.11 & 1 &  &  &  &  &  &  & -0.18 & & 0.01 & 1 &  & &  & \\
  NGC~1664 & both & 6.70 & 0.04 & 0.03 & 2 & 7.5 & 0.05 & 2 & 6.38 & 0.4 & 2 & -0.01 & & 0.01 & 1 &  & & & 2\\
  NGC~1798 & RGB & 2.55 & 1.05 & 0.33 & 10 & -2.0 & 1.7 & 9 & 2.66 & 0.82 & 4 & -0.18 & 0.02 & 0.01 & 10 & -0.18 & 0.02 & 9 & 3\\
  NGC~1817 & RGB & 66.20 & & 0.04 & 1 & 65.31 & 0.09 & 31 & 66.07 & 1.22 & 40 & -0.08 & & 0.01 & 1 & -0.08 & 0.02 & 5 & 2,4\\
  NGC~1857 & both & -1.00 & 5.68 & 1.80 & 10 &  &  &  & 0.49 & 0.76 & 6 & -0.12 & 0.01 & 0.01 & 2 &  &  &  & \\
  NGC~188 & RGB & -41.82 & 0.71 & 0.15 & 21 & -42.35 & 0.05 & 472 & -41.7 & 0.96 & 27 & 0.13 & 0.05 & 0.01 & 21 & 0.11 & 0.04 & 8 & 2,10\\
  NGC~1907 & RGB & 2.56 & 0.18 & 0.10 & 3 & 2.3 & 0.5 & 5 & 2.78 & 0.69 & 6 & -0.05 & 0.01 & 0.01 & 3 & -0.06 & 0.03 & 5 & 4\\
  NGC~1912 & RGB & -0.65 & 0.71 & 0.50 & 2 & 1.0 & 0.58 & 1 & -1.49 & 3.57 & 3 & -0.07 & 0.02 & 0.01 & 2 & -0.1 & 0.2 & 1 & 1\\
  NGC~2158 & RGB & 27.37 & 1.76 & 0.41 & 18 & 26.9 & 1.9 & 27 & 26.65 & 2.4 & 12 & -0.15 & 0.03 & 0.01 & 18 & -0.15 & 0.03 & 18 & 2,3\\
  NGC~2168 & MS & -6.37 & 0.54 & 0.14 & 15 & -8.17 & 0.22 & 3 & -7.9 & 0.75 & 14 & -0.13 & 0.07 & 0.02 & 11 &  & & & 2\\
  NGC~2183 & MS & 23.99 & 13.67 & 6.11 & 5 &  &  &  & 27.11 & 12.2 & 2 & -0.08 & 0.08 & 0.06 & 2 &  & &  & \\
  NGC~2232 & MS & 25.77 & & 0.11 & 1 & 26.6 & 0.77 & 4 & 25.35 & 3.39 & 16 & 0.04 & & 0.014 & 1 & 0.22 & 0.09 & 5 & 2,13\\
  NGC~2243 & RGB & 60.11 &  & 0.3 & 1 & 59.84 & 0.41 & 2 & 59.63 & 1.06 & 4 & -0.41 & & 0.01 & 1 & -0.38 & 0.04 & 16 & 2,7\\
  NGC~2244 & MS & 33.61 & 1.21 & 0.70 & 3 & 30.67 & 2.3 & 12 & 75.18 & 3.14 & 3 & -0.23 & 0.09 & 0.06 & 2 &  & & & 2\\
  NGC~2264 & MS & 21.65 & 1.78 & 0.35 & 25 & 20.7 & 1.53 & 10 & 20.22 & 8.51 & 2 &  &  &  &  &  &  &  & 1\\
  NGC~2304 & RGB & 39.49 & 9.92 & 4.05 & 6 & 43.4 & 0.35 & 2 & 53.29 & 6.72 & 2 & -0.09 & 0.09 & 0.04 & 6 &  & & & 1\\
  NGC~2318 & RGB & 8.26 & & 0.03 & 1 &  &  &  & 8.66 & 0.06 & 2 & 0.01 & & 0.01 & 1 &  & &  & \\
  NGC~2324 & RGB & 43.95 & 3.36 & 1.37 & 6 & 41.81 & 0.25 & 8 & 41.76 & 2.2 & 7 & -0.15 & 0.05 & 0.02 & 6 & -0.22 & 0.07 & 2 & 1,10\\
  NGC~2355 & RGB & 35.28 & & 0.03 & 1 & 35.1 & 0.14 & 9 & 36.85 & 1.72 & 6 & -0.11 & & 0.01 & 1 & -0.05 & 0.08 & 3 & 1,10\\
  NGC~2420 & RGB & 74.26 & 0.25 & 0.06 & 16 & 73.57 & 0.15 & 18 & 74.22 & 0.93 & 14 & -0.12 & 0.02 & 0.01 & 16 & -0.1 & 0.04 & 7 & 2,4\\
  NGC~2682 & both & 33.92 & 1.32 & 0.10 & 179 & 33.62 & 0.08 & 148 & 33.8 & 1.06 & 64 & 0.02 & 0.07 & 0.01 & 174 & 0.0 & 0.06 & 52 & 2,10\\
  NGC~366 & MS & -0.09 & 21.64 & 10.82 & 4 &  &  &  & 83.14 & 15.82 & 1 &  &  &  &  &  &  &  &  \\
  NGC~457 & MS & -11.57 & 0.88 & 0.44 & 4 & -32.7 & 3.47 & 7 &  &  &  &  &  &  &  &  &  &  & 1\\
  NGC~6469 & MS & -3.32 & & 0.13 & 1 &  &  &  &  &  &  &  &  &  &  &  &  &  & \\
  NGC~6494 & MS & -31.73 & & 0.01 & 1 & -8.18 & 0.09 & 4 & -8.37 & 2.09 & 9 &  &  &  &  &  &  &  & 2\\
  NGC~6531 & MS & -18.22 & 1.85 & 1.31 & 2 & -14.7 & 5.12 & 9 &  &  &  & 0.08 & & 0.01 & 1 &  & & & 1\\
  NGC~6649 & MS & 1.54 & 4.53 & 1.85 & 6 & -8.59 & 0.2 & 4 & -8.87 & 0.92 & 4 &  &  &  &  &  &  &  & 2\\
  NGC~6705 & RGB & 35.34 & 1.16 & 0.34 & 12 & 35.08 & 0.32 & 15 & 36.01 & 1.6 & 16 & 0.16 & 0.03 & 0.01 & 12 & 0.17 & 0.04 & 8 & 2,4\\
  NGC~6791 & RGB & -47.46 & 1.08 & 0.18 & 37 & -47.4 & 0.13 & 193 & -45.85 & 1.64 & 8 & 0.40 & 0.07 & 0.01 & 35 & 0.42 & 0.03 & 31 & 2,3\\
  NGC~6811 & both & 7.62 & 0.46 & 0.14 & 10 & 6.68 & 0.08 & 5 & 7.7 & 0.43 & 10 & -0.01 & 0.04 & 0.01 & 9 & -0.01 & 0.02 & 4 & 2,3\\
  NGC~6819 & RGB & 2.78 & 1.17 & 0.17 & 45 & 2.3 & 0.04 & 537 & 3.31 & 1.93 & 48 & 0.10 & 0.04 & 0.01 & 44 & 0.09 & 0.03 & 6 & 2,4\\
  NGC~6866 & RGB & 12.55 & 0.67 & 0.39 & 3 & 13.68 & 0.06 & 2 & 12.83 & 0.86 & 3 & 0.04 & 0.02 & 0.011 & 3 &  & & & 2\\
  NGC~6913 & MS & -17.42 & & 3.04 & 1 & -19.6 & 1.9 & 6 &  &  &  &  &  &  &  &  &  &  & 2\\
  NGC~7058 & MS & -19.58 & 0.15 & 0.07 & 4 &  &  &  & -16.78 & 3.53 & 5 & 0.12 & 0.04 & 0.02 & 4 &  & &  & \\
  NGC~7062 & RGB & -22.26 &  & 0.14 & 1 &  &  &  & -22.05 & 3.35 & 5 & 0.04 & & 0.01 & 1 &  & &  & \\
  NGC~7086 & MS & -20.14 & 6.87 & 2.43 & 8 & -18.95 & 0.76 & 1 & -22.34 & 4.4 & 4 &  &  &  &  &  &  &  & 2\\
  NGC~752 & MS & 6.13 & 0.90 & 0.52 & 3 & 5.54 & 0.14 & 54 & 5.69 & 0.88 & 94 & 0.01 & & 0.01 & 1 & 0.02 & 0.02 & 7 & 2,4\\
  NGC~7788 & MS & -7.28 & & 0.80 & 1 & -24.0 & 5.2 & 1 &  &  &  &  &  &  &  &  &  &  & 2\\
  NGC~7789 & both & -54.49 & 1.22 & 0.25 & 23 & -54.7 & 0.13 & 88 & -54.15 & 1.53 & 153 & 0.01 & 0.09 & 0.02 & 21 & 0.06 & 0.05 & 7 & 1,4\\
  Roslund~3 & MS & -12.91 & & 0.21 & 1 & -2.6 & 0.32 & 13 & -21.67 & 1.18 & 1 &  &  &  &  &  &  &  & 1\\
  Roslund~6 & MS & -6.09 & 8.24 & 4.12 & 4 &  &  &  & -12.89 & 5.92 & 51 &  &  &  &  &  &  &  &  \\
  RSG~7 & MS & -6.58 & 4.09 & 2.05 & 4 &  &  &  & -11.51 & 4.28 & 14 & 0.10 &  & 0.01 & 1 &  & &  & \\
  RSG~8 & MS & 2.47 & 4.49 & 2.01 & 5 &  &  &  & -9.64 & 5.75 & 16 &  &  &  &  &  &  &  & \\
  Ruprecht~148 & MS & 15.49 & & 0.03 & 1 &  &  &  &  &  &  &  &  &  &  &  &  &  & \\
  SAI~16 & RGB & -66.10 & 0.80 & 0.57 & 2 &  &  &  & -65.59 & 0.4 & 1 & -0.18 & 0.01 & 0.01 & 2 &  & &  & \\
  Stock~10 & MS & -16.16 & 4.53 & 2.03 & 5 & -8.16 & 10.0 & 1 & -11.67 & 2.08 & 16 &  &  &  &  &  &  &  & 2\\
  Stock~1 & MS & -10.22 & & 0.47 & 1 &  &  &  & -19.51 & 2.91 & 30 &  &  &  &  &  &  &  & \\
  Stock~2 & MS & 7.08 & & 0.09 & 1 & 1.8 & 1.82 & 27 & 8.19 & 1.77 & 183 &  &  &  &  &  &  &  & 1\\
  Stock~4 & MS & 13.08 & & 0.86 & 1 &  &  &  &  &  &  &  &  &  &  &  &  &  & \\
  Stock~7 & MS & 1.18 & & 0.01 & 1 &  &  &  & 4.79 & 0.09 & 2 &  &  &  &  &  &  &  & \\
  Teutsch~10 & RGB & 3.94 & & 0.08 & 1 &  &  &  & 1.72 & 5.5 & 2 & -0.30 & & 0.01 & 1 &  & &  & \\
  Teutsch~12 & both & 51.18 & 2.32 & 1.04 & 5 &  &  &  &  &  &  & -0.14 & 0.02 & 0.01 & 4 &  & &  & \\
  Teutsch~1 & MS & 16.29 & & 0.07 & 1 &  &  &  &  &  &  &  &  &  &  &  &  &  & \\
  Teutsch~51 & RGB & 17.13 & 0.59 & 0.26 & 5 & 17.0 & 1.4 & 5 &  &  &  & -0.28 & 0.04 & 0.02 & 5 & -0.28 & 0.03 & 5 & 3\\
  Teutsch~7 & RGB & 5.05 & 0.8 & 0.57 & 2 &  &  &  & 6.49 & 3.73 & 1 & 0.13 & 0.06 & 0.04 & 2 &  & &  & \\
  Tombaugh~4 & RGB & -89.84 & & 0.36 & 1 &  &  &  &  &  &  & -0.47 &  & 0.01 & 1 &  & &  & \\
  Trumpler~26 & RGB & -8.10 & 0.15 & 0.11 & 2 & -9.6 & 0.16 & 2 & -7.59 & 0.48 & 4 & 0.278 & 0.05 & 0.03 & 2 &  &  & & 2\\
  Trumpler~2 & MS & -5.15 & 9.03 & 4.04 & 5 & -27.5 & 11.5 & 2 & -4.12 & 0.28 & 11 &  &  &  &  &  &  &  & 7 \\
  Trumpler~3 & MS & 18.52 & 6.20 & 3.10 & 4 & -8.58 & 1.12 & 2 & -19.46 & 3.59 & 5 & -0.22 & & 0.01 & 1 &  & &  & 2\\
  Trumpler~5 & RGB & 49.99 & 1.34 & 0.47 & 8 & 49.46 & 1.84 & 3 & 51.33 & 1.4 & 16 & -0.36 & 0.02 & 0.01 & 8 & -0.4 & 0.01 & 3 & 11\\
\hline
\end{longtable}
\twocolumn
\end{landscape}
}

\end{document}